\documentclass[10pt,letterpaper,journal]{IEEEtran}

\usepackage{tikz}
\tikzstyle{arrow} = [thick,->,>=stealth]
\usetikzlibrary{decorations.pathreplacing}
\usepackage{pgfplots}
\pgfplotsset{compat=1.18}
\usepackage[cmex10]{amsmath}
\usepackage{amsthm}
\usepackage{amssymb,amsfonts,dsfont,bm,mathrsfs}
\usepackage{siunitx}
\usepackage[bookmarksopen=true,pagebackref=true]{hyperref}
\usepackage{cleveref}
\usepackage[noadjust]{cite}
\usepackage{graphicx}
\allowdisplaybreaks[2]
\usepackage{stfloats}
\usepackage{enumerate}
\usepackage{makecell}
\usepackage{multirow}
\usepackage{balance}
\usepackage{algorithmic}
\usepackage{algorithm}
\renewcommand{\algorithmicrequire}{\textbf{Input:}} 
\renewcommand{\algorithmicensure}{\textbf{Output:}} 
\makeatletter
\newcommand{\removelatexerror}{\let\@latex@error\@gobble}
\makeatother

\usepackage[right = 1.02in, left = 1.02in,top=0.96in,bottom=1.08in]{geometry}
{
\theoremstyle{plain}
\newtheorem{theorem}{Theorem}

\newtheorem{lemma}{Lemma}
\newtheorem{corollary}{Corollary}

\theoremstyle{definition}
\newtheorem{definition}{Definition}

\theoremstyle{remark}
\newtheorem{remark}{Remark}
}

\begin{document}

\title{Computation of a Unified Graph-Based Rate Optimization Problem} 

%

\author{Deheng Yuan, Tao Guo, {\it Member, IEEE},
Zhongyi Huang and Shi Jin, {\it Fellow, IEEE}%
\thanks{This paper was presented in part at the 2022 IEEE Information Theory Workshop.}%
\thanks{Deheng Yuan and Zhongyi Huang are with the Department of Mathematical Sciences, Tsinghua University, Beijing 100084, China (Emails: ydh22@mails.tsinghua.edu.cn, zhongyih@tsinghua.edu.cn).}
\thanks{Tao Guo is with the School of Cyber Science and Engineering, Southeast University, Nanjing 210096, China (Email: taoguo@seu.edu.cn).}
\thanks{Shi Jin is with the National Mobile Communications Research Laboratory, Southeast University, Nanjing, 210096, China (Email: jinshi@seu.edu.cn).}%
}

\maketitle



\ifCLASSOPTIONpeerreview
\begin{center} \bfseries EDICS Category: 3-BBND \end{center}
\fi
%

\begin{abstract}
We define a graph-based rate optimization problem and consider its computation, which provides a unified approach to the computation of various theoretical limits, including the (conditional) graph entropy, rate-distortion functions and capacity-cost functions with side information.
Compared with their classical counterparts, theoretical limits with side information are much more difficult to compute since their characterizations as optimization problems have larger and more complex feasible regions.
Following the unified approach, we develop effective methods to resolve the difficulty.
On the theoretical side, we derive graph characterizations for rate-distortion and capacity-cost functions with side information and simplify the characterizations in special cases by reducing the number of decision variables. 
On the computational side, we design an efficient alternating minimization algorithm for the graph-based problem, which deals with the inequality constraint by a flexible multiplier update strategy. Moreover, simplified graph characterizations are exploited and
deflation techniques are introduced, so that the computing time is greatly reduced. Theoretical analysis shows that the algorithm converges to an optimal solution. By numerical experiments, the accuracy and efficiency of the algorithm are illustrated and its significant  advantage over existing methods is demonstrated. 
\end{abstract}


\IEEEpeerreviewmaketitle

\section{Introduction}
\IEEEPARstart{T}{he} source coding and channel coding problems with side information are fundamental problems studied in information theory, e.g.,~\cite{Shannon1958,WynerZiv1976,Gelfand1980,Weissman2006,EIGamal2011,Cover2002,Barron2003,Pradhan2003}. 
Important special cases include the Wyner-Ziv lossy compression problem~\cite{WynerZiv1976} and the Gelfand-Pinsker channel problem~\cite{Gelfand1980}.
As an extension of the Wyner-Ziv problem, a lossy function computing problem with decoder side information was studied in~\cite{Yamamoto1982}. 
Theoretical limits for the problems are described by the rate-distortion and the capacity-cost functions. These functions reflect the fundamental trade-offs between communication resources and other constraints, such as the quality of service and input costs.
However, they were characterized by optimization problems that have much larger and more complex feasible regions,
{ which are described} in terms of auxiliary random variables and implicit reconstruction functions.
Hence the optimization problems is much more difficulty to solve, compared with their classical counterparts without side information.

For some of the function computing problems, graph characterizations are developed as effective tools for the computation of the corresponding optimal rates.
Orlitsky and Roche~\cite{Orlitsky2001} derived a graph-based characterization for the problem of lossless computing with decoder side information. 
They extended tools of {\it graph entropy} and {\it characteristic graph}, which were first introduced by K\"{o}rner~\cite{Korner1973} and Witsenhausen~\cite{Witsenhausen1976} for solving zero-error coding problems.
The auxiliary random variable involved therein is explicitly represented by the independent set of a characteristic graph.
The graph entropy approach in~\cite{Orlitsky2001} was generalized to the lossy computing problem.
Some work~\cite{Doshi2007,Doshi2010} characterized an achievable rate by defining the $D$-characteristic graph. 
Other work~\cite{Basu2020,Basu2022} generalized the independent sets to hyperedges and defined an $\epsilon$-characteristic hypergraph.
The rate-distortion function was characterized for a limited class of so-called maximal distortion measure which is defined as an indicator function. 
However, existing graph-based characterizations either led to an achievable but suboptimal rate~\cite{Doshi2007,Doshi2010} or failed to cope with general distortion measures~\cite{Basu2020,Basu2022} for the lossy computing problem.

{ In this work, we show that a wider class of rate optimization problems with side information can be unified into a single graph-based problem, by developing graph-based characterizations for them.
This motivates us to consider the computation of the unified problem, rather than each of these problems separately. 
Consequently, all these problems are immediately solved as special cases.}

{
\subsection{Previous Methods}
}
For the computation of classical rate-distortion and channel capacity problems, the Blahut-Arimoto (BA) iterative algorithm was first proposed by~\cite{Blahut1972,Arimoto1972} and has been thoroughly analyzed since then.
Based on that, many methods were proposed to increase the convergence speed~\cite{Matz2004,Huang2005,Yu2010,Sutter2015,Chandrasekaran2016,Nakagawa2017,Nakagawa2021}.
Moreover, several recent works aimed at improving the flexibility of BA algorithm~\cite{Wu2022,Hayashi2023,Chen2023}.
Specifically, the BA algorithm cannot compute the classical rate-distortion function for a fixed distortion criterion directly, because the Lagrange multiplier associated with the distortion constraint is a real input of the algorithm. 
The algorithm fixes the Lagrange multiplier during iterations, but the given fixed distortion criterion is not satisfied. 
To overcome the above weakness, convergent algorithms that can compute the classical rate-distortion function for the fixed distortion criterion were designed in~\cite{Hayashi2023,Chen2023}. This is achieved by updating the multiplier through a one-dimensional root-finding step 
in each iteration.

For the computation of rate optimization problems with side information, the difficulty mainly arises from { the} complexity of the feasible region represented in terms of auxiliary random variables and implicit reconstruction functions. 
It was not resolved by methods for the problems without side information~\cite{Matz2004,Huang2005,Yu2010,Sutter2015,Chandrasekaran2016,Nakagawa2017,Nakagawa2021,Wu2022,Hayashi2023,Chen2023} or tools in the broad optimization literature.
Efficient computation methods need to take advantage of more specific structures of the problems.
For the specific setting of lossless computing, graph characterizations in~\cite{Orlitsky2001} can be used for numerical computation and a BA type algorithm was developed by~\cite{Harangi2022}.
But as we have noted, for more general settings like the lossy computing problem, graph characterizations have not been developed and numerical computation is more difficult. 

To solve the lossy compression and capacity-cost problems with side information, some work~\cite{Cheng2005} traversed the reconstruction functions and optimized the distribution of the auxiliary random variable for each reconstruction function by a BA type algorithm. 
Other work~\cite{Willems1983,Dupuis2004} exploited a technique called the Shannon method that characterizes the auxiliary random variable as a function and eliminates the reconstruction function, and then designed a BA type algorithm to optimize the distribution of the random variable.
However, in both ways the computational cost increased sharply with the size of side information, since either the reconstruction functions was traversed~\cite{Cheng2005} or the alphabet of the auxiliary random variables was expanded~\cite{Willems1983,Dupuis2004}.
No effective methods were developed to mitigate the difficulty. 
Also, algorithms for the rate-distortion problem of computing a general function or with causal side information have not been designed.
Moreover, similar to the discussion for the classical counterparts~\cite{Wu2022,Hayashi2023,Chen2023}, 
the BA type algorithm suffers from the inconvenience { of searching} for the Lagrange multiplier, which incurs high computational complexity.

{
\subsection{Our Contributions}
Following the unified approach,} our contributions are summarized as follows:



{1)} We reduce the number of decision variables and simplify the problem in special cases, which helps with the computation both analytically and numerically.
Specifically, for the rate-distortion function of the lossy computing problem, we obtain a graph characterization for the general case. 
Our construction by a bipartite graph (or equivalently, a multi-hypergraph in our previous work~\cite{Yuan2022})  generalizes existing constructions for various special cases.  
Specifically, the independent sets in the characteristic graph in~\cite{Orlitsky2001} and the hyperedges in the $\epsilon$-characteristic hypergraph in~\cite{Basu2020} are generalized to vertices in the right partition of our characteristic bipartite graph.
For special cases of the lossless or approximate computing problems, the bipartite graph construction specialize to the graph or hypergraph 
construction and the characterizations in~\cite{Orlitsky2001,Basu2020} are recovered.
For the capacity-cost function with two-sided information, the graph characterizations are analogously developed, which provides a new view of the channel problem, dual to its source counterpart.

{2)} We design an alternating minimization algorithm to numerically solve the unified problem. 
Our algorithm copes with the inequality constraint in the problem directly, by designing flexible updating strategies for the Lagrange multiplier in the alternating minimization process. 
{With the multiplier updating strategy, our algorithm can compute the rate-distortion (capacity-cost) function for a given distortion (cost) criterion.
This resolves the inconvenience of searching for the multiplier suffered by the BA type algorithms in~\cite{Willems1983,Dupuis2004,Cheng2005}.}
The solutions generated by the algorithm are proved to converge to an optimal solution.
Furthermore, an $O(\frac{1}{n})$ convergence for the optimal value can be shown.

{3)}  We exploit simplified graph characterizations and develop  deflation techniques to accelerate the algorithm. 
Taking advantages of the graph characterizations, the number of decision variables can be exponentially reduced as the problems have specific structures.
Considering the sparsity of solutions, the number of decision variables can be further reduced {by performing deflation techniques during iterations}, 
which greatly reduces the computing time. 
{Both acceleration methods  effectively reduce the complexity of the feasible region in rate optimization problems with side information, which was not resolved by previous methods
for problems without side information~\cite{Wu2022,Hayashi2023,Chen2023}.
Moreover, these  effective methods overlooked  by~\cite{Willems1983,Dupuis2004,Cheng2005,Harangi2022}} can also accelerate the traditional BA type algorithms in~\cite{Willems1983,Dupuis2004,Harangi2022} and hence the computation of the whole rate-distortion (capacity-cost) curves.

{
4) The accuracy and efficiency of our algorithms are illustrated by numerical experiments.
Specialized to the computation of rate-distortion and capacity-cost functions with side information,
our algorithm is significantly faster than existing algorithms~\cite{Cheng2005,Willems1983,Dupuis2004}, even in hundreds of times if the same accuracy is achieved. 
The advantage
of our algorithm becomes more remarkable as the size
of the problem gets larger. As a result, our algorithm
can compute relatively large problems 
that existing algorithms fail to solve in a reasonable
time.}



\subsection*{Notations}
Denote a discrete random variable by a capital letter and its finite alphabet by the corresponding calligraphic letter, e.g., $V\in\mathcal{V}$ and $\hat{Z}\in\hat{\mathcal{Z}}$. 
Let $\mathscr{G}=(\mathscr{V},\mathscr{E})$ be a simple graph with the vertex set~$\mathscr{V}$ and edge set~$\mathscr{E}$. It is a {\it bipartite graph}~\cite{Bondy2008} if $\mathscr{V}$ can be split into disjoint sets $ \mathscr{V}_1$ and $\mathscr{V}_2$ so that each edge in $\mathscr{E}$ connects two vertices in $\mathscr{V}_1$ and $\mathscr{V}_2$, respectively. Denote such a bipartite graph by $\mathscr{G}[\mathscr{V}_1,\mathscr{V}_2, \mathscr{E}]$.
It is called {\it complete} if every vertex in $\mathscr{V}_1$ is joined to all vertices in~$\mathscr{V}_2$. 
Let $\omega : \mathscr{E} \to [0,1]$ be a weight function. 
The graph $\mathscr{G}[\mathscr{V}_1,\mathscr{V}_2, \mathscr{E}]$ associated with the weight $\omega$ is called a {\it weighted bipartite graph}, denoted by $\big(\mathscr{G}[\mathscr{V}_1,\mathscr{V}_2,\mathscr{E}],\omega\big)$.
\section{The Unified Graph-Based Optimization Problem}
\label{sec:pro}
Let $G[\mathcal{V},\mathcal{U},\mathcal{E}]$ be a bipartite graph and $\mathcal{W}$ be a finite set. 
Denote by $\mathcal{E}^u \subseteq \mathcal{V}$ and $\mathcal{E}_v \subseteq \mathcal{U}$ the set of vertices adjacent to $u$ and $v$, respectively.
Assume that $\mathcal{E}_v \neq \emptyset$ for any $v \in \mathcal{V}$. Let $\bm{p}_{U|V} = (p(u|v))_{u \in \mathcal{U}, v \in \mathcal{V}}$,
and we abbreviate it by $\bm{p}$ without ambiguity.
Given $p(v)$, $p(w|v,u)$, and a loss function $l: \mathcal{E} \times \mathcal{W} \to [0, \infty)$, we consider the  unified graph-based optimization problem as follows, 
\begin{subequations}\label{p0}
\begin{align}
\min_{\bm{p}}\  &I(U;V)-I(U;W),\label{p0a}
\\
\mathrm{s.t.}\  & \mathbb{P}[(V,U) \in \mathcal{E}] = 1, \label{p0b}
\\
&\mathbb{E}[l(V,U,W)] \leq L.\label{p0c}
\end{align}
\end{subequations}
Note that $p(w|v,u)$ is defined as $0$ for $(v,u) \notin \mathcal{E}$. Then we define the {\it rate-loss function} to be the optimal value of the problem~\eqref{p0}, which is denoted by $T(L)$.

Next we write the graph-based problem~\eqref{p0} into a more succinct form.
Denote the objective function $I(U;V)-I(U;W)$ by $\mathcal{O}(\bm{p})$. 
We can verify that the problem~\eqref{p0} depends on the loss function $l$ only through
 \begin{equation}
 \label{eq:defltilde}
\tilde{l}(v,u) \triangleq \sum_{w'} p(w'|u,v)l(v,u,w').
 \end{equation}
We denote by $\mathrm{supp}(\bm{p}) \triangleq \{(v,u)| p(u|v)>0\}$ the support of $p(u|v)$.
We call $G[\mathcal{V},\mathcal{U},\mathcal{E}]$ the {\it characteristic bipartite graph}  
for the problem~\eqref{p0}.
Each feasible solution $\bm{p}$ of \eqref{p0} naturally corresponds to a weight $\omega_{\bm{p}}$ on $G[\mathcal{V},\mathcal{U},\mathcal{E}]$ with $\omega_{\bm{p}}(v,u) = p(u|v)$. 
For any {\it subgraph} $G[\mathcal{V},\mathcal{H},\mathcal{F}]$ of $G[\mathcal{V},\mathcal{U},\mathcal{E}]$ (which satisfies $\mathcal{H} \subseteq \mathcal{U}$ and $\mathcal{F} \subseteq \mathcal{E} \cap \mathcal{V} \times \mathcal{H}$) and any $\omega_{\bm{p}}$, we say $\omega_{\bm{p}}$ is 
 {\it feasible} on $G[\mathcal{V},\mathcal{H},\mathcal{F}]$ if $\omega_{\bm{p}}$ is a weight on $G[\mathcal{V},\mathcal{H},\mathcal{F}]$, or equivalently, $\mathrm{supp}(\bm{p}) \subseteq \mathcal{F}$.
Then let 
\begin{equation}
    \Omega(\mathcal{H},\mathcal{F}) = \big\{\omega| \omega \text{ is feasible on } G[\mathcal{V},\mathcal{U},\mathcal{F}]\big\}.
\end{equation} 

We immediately have 
\begin{equation}
\label{lem:pgraph}
T(L) = \min_{\omega_{\bm{p}} \in \Omega(\mathcal{U},\mathcal{E})} \mathcal{O}(\bm{p}).
\end{equation}

Without loss of generality, we can assume that for any $v$,
\begin{equation}\label{positivepv}
    p(v)>0,
\end{equation}
and for any $w$,
\begin{equation}\label{positivepw}
    \exists (v,u) \in \mathcal{E},\  \mathrm{s.t. }\ p(w|v,u)>0.
\end{equation}
Otherwise, if the assumptions of \eqref{positivepv} and \eqref{positivepw} are not satisfied, we can just eliminate such $v$ and $w$.

In the rest of this section, we give several examples of the graph-based problem~\eqref{p0}. They are optimization problems induced by specific source and channel coding problems. 

\subsection{Graph Entropy}

The first example is the graph entropy problem. The problem was revealed in the study of one source zero-error coding problem~\cite{Korner1973} and its independent values were discovered afterwards~\cite{Simonyi1995}.

Let $\mathcal{G}_0 = (\mathcal{V}_0,\mathcal{E}_0)$ be a graph and $V_0$ be a random variable over $\mathcal{V}_0$.
Let $\Gamma(\mathcal{G}_0)$ to be the collection of independent sets of $\mathcal{G}_0$.
The graph entropy of $V_0$ is defined as 
\begin{equation}
\label{eq:graphentropy}
H_{\mathcal{G}_0}(V_0) = \min_{p(u|v_0): V_0 \in U \in \Gamma(\mathcal{G}_0)} I(U;V_0).
\end{equation}
In the graph-based problem~\eqref{p0}, let $V = V_0$, $U = U_0$, $W = \emptyset$, $\mathcal{V} = \mathcal{V}_0$, $\mathcal{U} = \Gamma(\mathcal{G}_0)$ and construct $\mathcal{E} = \{(v_0,u_0)| v_0 \in u_0 \in \Gamma(\mathcal{G}_0)\}$, then the constraint $V_0 \in U \in \Gamma(\mathcal{G}_0)$ in~\eqref{eq:graphentropy} is equivalent to~\eqref{p0b}. 
By directly discarding the additional loss constraint~\eqref{p0c}, the graph entropy problem~\eqref{eq:graphentropy} is a special case of the graph-based problem~\eqref{p0}. 

\subsection{Conditional Graph Entropy}
The following conditional graph entropy problem generalizes the graph entropy problem, and was used in~\cite{Orlitsky2001} to characterize the optimal rate for lossless computing problem.

Let $\mathcal{G}_0 = (\mathcal{V}_0,\mathcal{E}_0)$ be a graph, $(V_0, W_0)$ be a pair of random variables and $V_0 \in \mathcal{V}_0$.
Let $\Gamma(\mathcal{G}_0)$ to be the collection of independent sets of $\mathcal{G}_0$.
The conditional graph entropy of $V_0$ given $W_0$ is defined as 
\begin{equation}
\label{eq:conditionalgraphentropy}
H_{\mathcal{G}_0}(V_0|W_0) = \min_{p(u|v_0): V_0 \in U \in \Gamma(\mathcal{G}_0)} I(U;V_0|W_0).
\end{equation}
In the graph based problem~\eqref{p0}, let $V = V_0$, $U = U_0$, $W = W_0$, $\mathcal{V} = \mathcal{V}_0$, $\mathcal{U} = \Gamma(\mathcal{V}_0)$, $p(w_0|v_0,u_0) = p(w_0|v_0)$ and construct $\mathcal{E} = \{(v_0,u_0)| v_0 \in u_0 \in \Gamma(\mathcal{G}_0)\}$, then the constraint $V_0 \in U \in \Gamma(\mathcal{G}_0)$ in~\eqref{eq:conditionalgraphentropy} is equivalent to~\eqref{p0b}. 
Discarding the additional loss constraint~\eqref{p0c} shows that the conditional graph entropy problem~\eqref{eq:conditionalgraphentropy} is a special case of the graph-based problem~\eqref{p0}.

Next we consider some examples that are not so direct. 
\subsection{Rate-Distortion Problems for Lossy Computing}
\label{eg:rdf}
Consider the rate-distortion problem for the  lossy computing problem with two-sided information. 
Let $(S_{1},S_{2}) \sim p(s_1,s_2)$ be discrete memoryless sources distributed over $\mathcal{S}_1\times\mathcal{S}_2$. 
Without loss of generality, assume $p(s_1) > 0$ and $p(s_2)>0$, $\forall s_1 \in \mathcal{S}_1$, $s_2 \in \mathcal{S}_2$.
The source message $S_1$ observed by the encoder has two parts $(S,\hat{S}_{1})$, with $S$ being the original source and $\hat{S}_1$ being the encoder side information. 
The other part $S_2$ of the side information is observed by the decoder.
The decoder needs to compute a function $f : \mathcal{S}_1 \times \mathcal{S}_2 \to \mathcal{Z}$ within a certain distortion. 
Denote $f(S_{1},S_{2})$ by $Z$. Let $d: \mathcal{Z} \times \hat{\mathcal{Z}} \to [0,\infty)$ be a distortion measure. 
Then the results of~\cite{Cover2002} can be adapted to show that the rate-distortion function can be written as
\begin{equation}\label{prdf}
R(D) = \min_{\substack{p(u|s_1):\  \exists g, \\  \mathbb{E}[d(f(S_1,S_2),g(U,S_2))] \leq D}}I(U;S_1)-I(U;S_2).
\end{equation}


 
	
	
	
	

Computation of the rate-distortion function by directly solving the optimization problem~\eqref{prdf} is difficult, since it needs to traverse the implicit reconstruction function $g$ and then optimize the auxiliary random variable $U$ for each fixed~$g$, which leads to high complexity (cf.~\cite{Cheng2005}). To circumvent the difficulty, we first transform the problem into an equivalent, special case of the graph-based problem~\eqref{p0}, and then develop effective methods for~\eqref{p0}.
The following lemma shows the corresponding rate-distortion problem is a special case of \eqref{p0} by assigning
\begin{subequations}
\label{eq:assrdf}
\begin{align}
&V = S_1,~U = (\hat{Z}_{s_2})_{s_2 \in \mathcal{S}_2},~W = S_2,
\\
&\mathcal{V} = \mathcal{S}_1,~\mathcal{U} = \hat{\mathcal{Z}}^{\mathcal{S}_2},~\mathcal{W} = \mathcal{S}_2,~\mathcal{E} = \mathcal{V} \times \mathcal{U},
\\
&l(s_1,(\hat{z}_{s_2})_{s_2 \in \mathcal{S}_2},s_2) = d(f(s_1,s_2),\hat{z}_{s_2}),~L = D,
\\
&p(s_2|u,s_1) = p(s_2|s_1), \label{eq:assrdfd}
\end{align}
\end{subequations}
where $(\hat{z}_{s_2})_{s_2 \in \mathcal{S}_2}$ is a length $|\mathcal{S}_2|$ vector indexed by $s_2 \in \mathcal{S}_2$ and $\hat{\mathcal{Z}}^{\mathcal{S}_2} = \big\{(\hat{z}_{s_2})_{s_2 \in \mathcal{S}_2}|\hat{z}_{s_2} \in \hat{\mathcal{Z}}, \forall s_2 \in \mathcal{S}_2\big\}$. The proof can be found in Appendix~B {of the complete version of the current work~\cite{Yuan2023}.} 
\begin{lemma}\label{lem:rdf}
The rate-distortion function for the lossy computing problem can be characterized by
\begin{equation}\label{eq:rdf}
R(D) 
= \min_{\substack{p(u|s_1):U = (\hat{Z}_{s_2})_{s_2 \in \mathcal{S}_2}, \\  \mathbb{E}[d(f(S_1,S_2),\hat{Z}_{S_2})] \leq D}}I(U;S_1)-I(U;S_2).
\end{equation}
\end{lemma}

\begin{remark}
For the rate-distortion function of the lossy computing problem, \Cref{lem:rdf} gives a graph characterization for the general case. 
Existing graph constructions, such as the hyperedges in the $\epsilon$-characteristic hypergraph in~\cite{Basu2020}, are only for the special case $D = 0$ and can not handle the general case here. There are two main reasons.
Firstly, the region of the feasible weights is significantly reduced for $D = 0$. 
The intuition behind is the ``zero effect" for $D=0$, i.e., for each edge $(s_1,u)$ with $p(u|s_1)>0$, the loss $\tilde{l}(s_1,u)$ induced by $(s_1,u)$ must be zero.
However, for $D>0$, even the edge $(s_1,u)$ inducing a loss larger than~$D$ are still possible, since the average loss is of final concern.
Secondly, for $D = 0$ the candidate recovery for each $u$ is chosen fully based on $\mathcal{E}^u$.
However, for $D>0$, it is necessary that the induced reconstruction may take different values in $\hat{\mathcal{Z}}$ for $u$ with the same $\mathcal{E}^u$, in order to achieve a smaller average distortion with a limited rate.
In~\Cref{subsec:mindist}, we show that our graph characterizations can specialize to the results in previous work~\cite{Orlitsky2001,Basu2020} for~$D = 0$. 
\end{remark}



\begin{remark}
Similar results for lossy compression ($f(x,y) = x$) without the encoder side information ($\hat{S}_1 = \emptyset$) were obtained using Shannon strategy in \cite{Willems1983,Dupuis2004}, which were subsumed by \Cref{lem:rdf} as a special case. 
Similarly,~\Cref{lem:ccf} 
 in~\Cref{eg:ccf} subsumes the special case considered in~\cite{Dupuis2004} without decoder side information ($S_2 = \emptyset$).
\end{remark}
\begin{remark}
Note that there is an equivalence between the bipartite graph and the multi-hypergraph defined in~\cite{Yuan2022}. Then~\Cref{lem:rdf} can also be written in terms of the multi-hypergraph. We adopt the bipartite graph approach here.
\end{remark}

\subsection{Capacity-Cost Problems with Two-Sided Information}
\label{eg:ccf}
Consider the channel coding problem with two-sided state information. 
Let $(\mathcal{X},p(y|x,s_1,s_2),\mathcal{Y},\mathcal{S}_1 \times \mathcal{S}_2)$ be a discrete-memoryless channel with state information $(S_{1},S_{2}) \sim p(s_1,s_2)$ distributed over $\mathcal{S}_1\times\mathcal{S}_2$.
We assume that $S_1$ and $S_2$ are respectively observed by the encoder and the decoder, and $p(s_1) > 0$, $\forall s_1 \in \mathcal{S}_1$.
Let $b: \mathcal{X} \times \mathcal{S}_1 \times \mathcal{S}_2 \to [0,\infty)$ be a cost measure depending on the input and the channel state. 
	

 
	


        
        

 
The following lemma is proved in Appendix~C in~\cite{Yuan2023}. 
\begin{subequations}
\label{eq:assccf}
\begin{align}
&V = S_1,~U = (X_{s_1'})_{s_1' \in \mathcal{S}_1},~W = (Y,S_2),
\\
&\mathcal{V} = \mathcal{S}_1,~\mathcal{U} = \mathcal{X}^{\mathcal{S}_1},~\mathcal{W} = \mathcal{Y} \times \mathcal{S}_2,~\mathcal{E} = \mathcal{V} \times \mathcal{U},
\\
&l(s_1,(x_{s_1'})_{s_1 '\in \mathcal{S}_1},(y,s_2)) = b(x_{s_1},s_1,s_2), L = B,
\\
&p(y,s_2|(x_{s_1'})_{s_1 '\in \mathcal{S}_1},s_1) = p(y|x_{s_1},s_1,s_2)p(s_2|s_1).
\label{eq:assccfd}
\end{align}
\end{subequations}

\begin{lemma}
\label{lem:ccf}
The capacity-cost function for the channel coding problem with state information can be alternatively characterized by
\begin{equation}\label{eq:ccf}
C(B) 
=  -\min_{\substack{p(u|s_1): U = (X_{s_1})_{s_1 \in \mathcal{S}_1},\\ \mathbb{E}[b(X_{S_1},S_1,S_2)] \leq B}}I(U;S_1)-I(U;Y,S_2). 
\end{equation}
\end{lemma}

{
\begin{remark}
Further consider the case where the number of side information is greater than two.
In this case, the side information obtained by the encoder and the decoder can be combined respectively.
Then the problem falls into the setting in this subsection and can be handled by our unified method.
For instance, suppose that the encoder side information is $S_{11}$ and $ S_{12}$, and the decoder side information is $S_{21}$, $S_{22}$, and $S_{23}$. 
By letting $S_1 = (S_{11},S_{12})$ and $S_{2} = (S_{21},S_{22},S_{23})$, the capacity-cost function is obtained by~\Cref{lem:ccf}, and becomes a special case of the unified problem in~\eqref{p0}.
Similar arguments also hold for the lossy computing problems in~\Cref{subsec:rdfside} and are not limited to problems discussed here.

\end{remark}}





\subsection{Rate-Distortion Problems with Causal Decoder Information}
\label{eg:crdf}
Consider the function computing problem with causal decoder side information. The setting is the same as~\Cref{eg:rdf}, except that the side information $S_2$ is causally known to the decoder (cf.~\cite{Weissman2006} and Section 11.2 in~\cite{EIGamal2011}).
The results in~\cite{Weissman2006} can be adapted to show that the rate-distortion function can be written as
\begin{equation}\label{pcrdf}
R(D) = \min_{\substack{p(u|s_1):\  \exists g, \\  \mathbb{E}[d(f(S_1,S_2),g(U,S_2))] \leq D}}I(U;S_1).
\end{equation}
By assigning
\begin{subequations}
\label{eq:asscrdf}
\begin{align}
&V = S_1,~U = (\hat{Z}_{s_2})_{s_2 \in \mathcal{S}_2},~W = \emptyset,
\\
&\mathcal{V} = \mathcal{S}_1,~\mathcal{U} = \hat{\mathcal{Z}}^{\mathcal{S}_2},~\mathcal{W} =  \emptyset,~\mathcal{E} = \mathcal{V} \times \mathcal{U},~L = D,
\\
&l(s_1,(\hat{z}_{s_2})_{s_2 \in \mathcal{S}_2}) = \sum_{s_2} p(s_2|s_1) d(f(s_1,s_2),\hat{z}_{s_2}),
\end{align}
\end{subequations}
then the problem~\eqref{pcrdf} is a special case of \eqref{p0}, summarized in the following lemma. 
\begin{lemma}\label{lem:crdf}
The rate-distortion function for the lossy computing problem with causal side information can be characterized by
\begin{equation}\label{eq:crdf}
R(D) 
= \min_{\substack{p(u|s_1):U = (\hat{Z}_{s_2})_{s_2 \in \mathcal{S}_2}, \\  \mathbb{E}[d(f(S_1,S_2),\hat{Z}_{S_2})] \leq D}}I(U;S_1).
\end{equation}
\end{lemma}
The proof is similar to that for~\Cref{lem:rdf} {but simpler. See Appendix~D in~\cite{Yuan2023} for details. }

\section{Properties of the Graph-Based Problem}
\label{sec:property}
The problem~\eqref{p0} is transformed into an equivalent form in~\Cref{subsec:equ} and the properties of the rate-loss function are investigated in~\Cref{subsec:los}, preparing for the designing of numerical computation algorithms.

\subsection{Properties and Equivalent Forms of the Graph-Based Problem}
\label{subsec:equ}

The problem~\eqref{p0} can be written as { an} equivalent form, which motivates our alternating minimization algorithms. To see this, we use $q(u|v)$ and $r(u|w)$ to replace $p(u|v)$ and $p(u|w)$. 
Then $\bm{q}$ and $\bm{r}$ are defined accordingly similar to $\bm{p}$.
Define
\begin{equation}
\label{eq:def-loss}
Loss(\bm{q}) = \sum_{(v,u)\in \mathcal{E}} p(v)q(u|v) \tilde{l}(v,u) {;}
\end{equation}
and the {\it generalized Kullback-Leibler (K-L) divergence} 
\begin{equation}
\label{eq:def-GD}
GD_{\mathcal{E}}(\bm{q}||\bm{r}) \!=\! \sum_{(v,u)\in \mathcal{E},w} \! p(v)q(u|v)p(w|u,v) \log \frac{q(u|v)}{r(u|w)}.
\end{equation}
Unlike the classical K-L divergence, $GD_{\mathcal{E}}(\bm{q}||\bm{r}) \geq 0$ does not always hold.

In light of~\eqref{eq:def-loss} and \eqref{eq:def-GD}, the equivalent form of~\eqref{p0} can be written as
\begin{subequations}\label{p2}
\begin{align}
\min_{\bm{q},\bm{r}}\  & GD_{\mathcal{E}}(\bm{q}||\bm{r}),\label{p2a}
\\
\mathrm{s.t.}\ &\mathrm{supp}(\bm{q}) \subseteq \mathcal{E}, \label{p2b} 
\\
&Loss(\bm{q}) \leq L.\label{p2c}
\end{align}
\end{subequations}




The problem~\eqref{p0} and its equivalent form~\eqref{p2} share fine properties given in the following lemma. The proof can be found in Appendix~{E} in~\cite{Yuan2023}. 

\begin{lemma}\label{lem:con opt}
The problems \eqref{p0} and  \eqref{p2} have the same optimal value $T(L)$. Moreover, they are both convex  problems.
\end{lemma}

\subsection{Properties of the Rate-Loss Function}
\label{subsec:los}
The  rate-loss function has many useful properties. 
The following lemma is proved in Appendix~{F} in~\cite{Yuan2023}. 

\begin{lemma}
\label{lemma:T(L)}
The rate-loss function $T(L)$ is non-increasing and convex in $L \in [0,\infty)$.
\end{lemma}

Then we identify different cases of problem~\eqref{p0}, which is useful for the numerical computation.
In order for that, define boundaries for the cases to be 
\begin{equation}
    L_{min} \triangleq \min_{\mathrm{supp}(\bm{q}) \subseteq \mathcal{E}} Loss(\bm{q}),
\end{equation}
\begin{equation}
    L_{max} \triangleq \mathop{\arg\min}_{L' \geq 0} T(L'),
\end{equation}
\begin{equation}
    L_{Max} \triangleq \max_{\mathrm{supp}(\bm{q}) \subseteq \mathcal{E}} Loss(\bm{q}).
\end{equation}
We can see  $L_{min} \leq L_{max} \leq L_{Max}$, and $L_{min}$ and $L_{Max}$ can be easily computed by
\begin{equation}\label{eq:L_min}
    L_{min} = \sum_{v} p(v)\min_{u:(v,u) \in \mathcal{E}} \left\{\tilde{l}(v,u)\right\},
\end{equation}
\begin{equation}\label{eq:L_Max}
    L_{Max} = \sum_{v} p(v)\max_{u:(v,u) \in \mathcal{E}} \left\{ \tilde{l}(v,u)\right\}.
\end{equation}
In contrast, $L_{max}$ does not have an explicit formula in general, except for special cases such as the lossy computing problem in \Cref{sec:rdf}. 

The cases identified by $L_{min}$, $L_{max}$ and $L_{Max}$ are listed in the following lemma, which can be easily derived from \Cref{lemma:T(L)} and we omit the detailed proof. 

\begin{lemma}\label{thm:classify0}
For the problem \eqref{p0}, we have the following.
\begin{enumerate}[i)]
\item For $0\leq L<L_{min}$, \eqref{p0} is infeasible, so $T(L) = \infty$.

\item For $L\geq L_{min}$, \eqref{p0} is feasible and $T(L_{min}) < \infty$.

\item For $L>L_{min}$,  {Slater's Constraint Qualification (SLCQ)\footnote{SLCQ is a condition on the convex optimization problem, under
which the Karush-Kuhn-Tucker (KKT) conditions are both necessary and sufficient for optimality.
It requires that there exists an feasible solution that lets the inequality constraints of the problem
hold with strict inequalities.
See Chapter 5 in~\cite{Boyd2004} for detailed discussions.}} is satisfied and  the Karush-Kuhn-Tucker (KKT) conditions are both necessary and sufficient for optimality.
 
\item $T(L)$ is continuous for $L \in [L_{min}, \infty)$.

 \item $T(L)$ is strictly decreasing for $L \in [L_{min},L_{max}]$. In this case, the optimal value of the problem \eqref{p0} is achieved when the equality in \eqref{p0c} holds.
 
\item For $ L \geq L_{Max}$, \eqref{p0c} is naturally satisfied and $T(L) = T(L_{Max})$.

\end{enumerate}
\end{lemma}

\begin{remark}
    The problem~\eqref{p2} shares similar properties as described in~\Cref{thm:classify0}.
\end{remark}

In Fig~\ref{fig:tlfunction}, the rate-loss function $T(L)$ is plotted and compared with two special cases, the rate-distortion function $R(D)$ in~\Cref{eg:rdf} and the capacity-cost function $C(B)$ in~\Cref{eg:ccf}.
In light of \Cref{thm:classify0}, numerical computations for the problem~\eqref{p0} can be done as follows. First compute $L_{min}$ and $L_{Max}$ by~\eqref{eq:L_min} and \eqref{eq:L_Max}.
By i) and ii), only $L \geq L_{min}$ is feasible.
By vi), $T(L) = T(L_{Max})$ for $L \geq L_{Max}$.
Then we confine to $L \in [L_{min},L_{Max}]$ in the sequel. 


\begin{figure}[!t]
\centering
\begin{tikzpicture}[scale = 0.6]
\tikzstyle{every node}=[font=\small,scale=0.7]
\begin{axis}[axis lines=middle,
            samples = 300,
            xtick = {0},
            ytick = {0},
            xmin = 0, xmax = 8,
            ymin = -0.5,ymax = 1.2,
            ]
    \addplot[domain = 1:5,blue,thick]{-ln(\x+1)+1.5};
    \addplot[domain = 5:7,blue,thick]{-ln(6)+1.5};
\end{axis}
\fill[color = blue] (0.85,4.43) circle (0.05);
\draw[dotted] (0.85,4.43)--(0,4.43);
\node at (-0.7,4.43) {$T(L_{min})$};
\draw (0,4.43)--(0.1,4.43);

\draw[dotted] (0.85,4.43)--(0.85,1.7);
\node at (0.85,1.4) {$L_{min}$};
\draw (0.85,1.7)--(0.85,1.8);

\draw[dotted] (4.3,0.7)--(0,0.7);
\node at (-0.7,0.7) {$T(L_{Max})$};
\draw (0,0.7)--(0.1,0.7);

\draw[dotted] (4.3,0.7)--(4.3,1.7);
\node at (4.3,2.0) {$L_{max}$};
\draw (4.3,1.7)--(4.3,1.8);

\draw[dotted] (5.4,0.7)--(5.4,1.7);
\node at (5.4,2.0) {$L_{Max}$};
\draw (5.4,1.7)--(5.4,1.8);

\node at (6.8,1.4) {$L$};
\node at (-0.5,5.6) {$T(L)$};

\draw[dotted,red,thick] (-1.5,-0.05) rectangle (7,6);
\end{tikzpicture}

\begin{tikzpicture}[scale = 1]
\tikzstyle{every node}=[font=\small,scale=1]
\draw [->,thick] (-1,0) to  (-1,-0.5);
\draw [->,thick] (1,0) to (1,-0.5);
\node at (0,-0.25) {Specialize};
\end{tikzpicture}

\begin{tikzpicture}[scale = 0.325]
\tikzstyle{every node}=[font=\small,scale=0.325]
\begin{axis}[axis lines=middle,
            samples = 300,
            xtick = {0},
            ytick = {0},
            xmin = 0, xmax = 8,
            ymin = 0,ymax = 1.2,
            ]
    \addplot[domain = 1:5,blue,thick]{-ln(\x+2)+ln(7)};
    \addplot[domain = 5:7,blue,thick]{0};
\end{axis}
\fill[color = blue] (0.85,4.06) circle (0.05);
\draw[dotted] (0.85,4.06)--(0,4.06);
\node at (-0.7,4.06) {$R(D_{min})$};
\draw (0,4.06)--(0.1,4.06);

\draw[dotted] (0.85,4.06)--(0.85,0);
\node at (0.85,-0.3) {$D_{min}$};
\draw (0.85,0)--(0.85,0.1);

\node at (4.3,-0.3) {$D_{max}$};
\draw (4.3,0)--(4.3,0.1);

\node at (6.8,-0.3) {$D$};
\node at (-0.5,5.6) {$R(D)$};
\draw[dotted,red,thick] (-1.5,-0.5) rectangle (7.5,6);
\end{tikzpicture}
\begin{tikzpicture}[scale = 0.325]
\tikzstyle{every node}=[font=\small,scale=0.325]
\begin{axis}[axis lines=middle,
            samples = 300,
            xtick = {0},
            ytick = {0},
            xmin = 0, xmax = 8,
            ymin = -2,ymax = 0.5,
            ]
    \addplot[domain = 1:5,blue,thick]{-ln(\x+0.5)+ln(1.3)};
    \addplot[domain = 5:7,blue,thick]{-ln(5.5)+ln(1.3)};
\end{axis}
\fill[color = blue] (0.85,4.25) circle (0.05);
\draw[dotted] (0.85,4.25)--(0,4.25);
\node at (-0.9,4.25) {$-C(B_{min})$};
\draw (0,4.25)--(0.1,4.25);

\draw[dotted] (0.85,4.25)--(0.85,4.55);
\node at (0.85,4.85) {$B_{min}$};
\draw (0.85,4.45)--(0.85,4.55);

\draw[dotted] (4.3,1.25)--(0,1.25);
\node at (-0.9,1.25) {$-C(B_{Max})$};
\draw (0,1.25)--(0.1,1.25);

\draw[dotted] (4.3,1.25)--(4.3,4.55);
\node at (4.3,4.85) {$B_{max}$};
\draw (4.3,4.45)--(4.3,4.55);

\draw[dotted] (5.4,1.25)--(5.4,4.55);
\node at (5.4,4.85) {$B_{Max}$};
\draw (5.4,4.45)--(5.4,4.55);

\node at (6.8,4.85) {$B$};
\node at (-0.65,5.6) {$-C(B)$};
\draw[dotted,red,thick] (-1.8,-0.5) rectangle (7.5,6);
\end{tikzpicture}
\caption{The rate-loss curve $T(L)$, compared with its special cases, the rate-distortion curve $R(D)$ in~\Cref{eg:rdf} (or the rate-distortion curve in~\Cref{eg:crdf}) and the capacity-cost curve $C(B)$ in~\Cref{eg:ccf}.}
\label{fig:tlfunction}
\vspace{-0.2cm}
\end{figure}

\section{Simplifying Graph Characterizations for Rate-Distortion and Capacity-Cost Functions}
\label{sec:rdfccf}
We simplify the graph characterization~\eqref{eq:rdf} and~\eqref{eq:ccf} for rate-distortion and capacity-cost problems with two-sided non-casual information in~\Cref{sec:rdf,sec:ccf}.
Further note that graph characterizations for problems with casual information (e.g.~\Cref{eg:crdf}) can be simplified as well, though details are not presented in this work due to the space limitation.

\subsection{Simplifying Graph Characterizations in~\eqref{eq:rdf}}
\label{sec:rdf}

\subsubsection{The Case that $S_1$ and $S_2$ have a G\'{a}cs-K\"{o}rner-Witsenhausen Common Part}
\label{subsec:rdfside}
First recall the following standard definition in~\cite{EIGamal2011}.
\begin{definition}[G\'{a}cs-K\"{o}rner-Witsenhausen common part~\cite{EIGamal2011}]
\label{def:GKW}
Let $(S_1,S_2)$ be a pair of discrete random variables. 
By relabeling the alphabets $\mathcal{S}_1$ and $\mathcal{S}_2$, we can arrange $p(s_1,s_2)$ in a block diagonal form, where there are at most $K$ nonzero blocks.
The {\it common part} of $S_1$ and $S_2$ is the random variable $S_0$ that takes value $k$ if $(S_1,S_2)$ is in block $k$, $k = 1,...,K$. In other words, there exists some function $g_1: \mathcal{V} \to \{1,...,K\}$, $g_2: \mathcal{V}' \to  \{1,...,K\}$ such that $S_0 = g_1(S_1) = g_2(S_2)$.
\end{definition}

Now suppose the two-sided information $S_1$ and $S_2$ have a G\'{a}cs-K\"{o}rner-Witsenhausen common part, denoted by $S_0 \in \{1,...,K\}$. Let $S_0 = g_1(S_1) = g_2(S_2)$ and $\mathcal{S}_{2k} =  g_2^{-1}(k)$, $k = 1,...,K$, {where $g_2^{-1}(k) = \{s_2 \in S_2| g_2(s_2) = k\}$ is the preimage of $k$ under the map~$g_2$}.  
Then we have the following theorem proved in Appendix~{G} in~\cite{Yuan2023}.

\begin{theorem}\label{thm:rdfci}
If $S_1$ and $S_2$ have a G\'{a}cs-K\"{o}rner-Witsenhausen common part, then
\begin{equation}\label{eq:rdfci}
R(D) = \min_{\substack{p(u|s_1):S_1 \in \mathcal{S}_1, U \in \mathcal{U}'_{S_1},\\  \mathbb{E}[d(f(S_1,S_2),\hat{Z}_{S_2})] \leq D}}I(U;S_1)-I(U;S_2),
\end{equation}
where $\mathcal{U}'_{S_1} = \hat{\mathcal{Z}}^{g_2^{-1}(g_1(S_1))}$.
\end{theorem}

Furthermore, if both $S_1$ and $S_2$ can be partitioned into two parts, we can simplify~\eqref{eq:rdfci} as follows.

\begin{corollary}\label{cor:rdfci}
If $S_1 = (S_0,S_1')$ and $S_2 = (S_0,S_2')$, then
\begin{equation}\label{eq:rdfcisim}
R(D) = \min_{\substack{p(u|s_1): U = (\hat{Z}_{s_2'})_{s_2' \in \mathcal{S}_2'},\\  \mathbb{E}[d(f(S_1,S_2),\hat{Z}_{S_2})] \leq D}}I(U;S_1)-I(U;S_2).
\end{equation}
\end{corollary}

Note that \eqref{eq:rdfcisim} is a special case of~\eqref{p0}. 
The alphabet $\mathcal{U}$ is $\hat{\mathcal{Z}}^{\mathcal{S}_2'}$, which is strictly smaller than $\hat{\mathcal{Z}}^{\mathcal{S}_2}$ given by \Cref{lem:rdf}. Hence the number of decision variables significantly decreases from  $|\mathcal{S}_1|\cdot |\hat{\mathcal{Z}}|^{|\mathcal{S}_2|}$ to $|\mathcal{S}_1|\cdot |\hat{\mathcal{Z}}|^{|\mathcal{S}_2'|}$.

\subsubsection{Minimum Distortion Case}\label{subsec:mindist}
Consider the minumum loss case that $L = L_{min}$ and we simplify the optimization problem in~\eqref{eq:rdf}. 
For any $e= (s_1,u) \in \mathcal{E}$,
recall that by~\eqref{eq:assrdfd} we have 
\begin{equation*}
    \tilde{l}(s_1,u) 
=\sum_{s_2}p(s_2|s_1)d(f(s_1,s_2),\hat{z}_{s_2}),
\end{equation*}
where $u = (\hat{z}_{s_2})_{s_2 \in \mathcal{S}_2}$. Then let $e \in \mathcal{E}_*$ if  
\begin{equation}
    \tilde{l}(s_1,u) = \min_{u' \in \mathcal{E}_{s_1}}\left\{\tilde{l}(s_1,u')\right\}.
\end{equation}
By deleting vertices in $\mathcal{U}$ that are not adjacent to any edges in $\mathcal{E}_*$, we obtain $\mathcal{U}_* =\cup_{v \in \mathcal{V}}(\mathcal{E}_{*})_v$.
Recall $\mathcal{E}_*^u = \{v \in \mathcal{S}_1| (v,u) \in \mathcal{E}_*\}$, then we define
\begin{equation}
\label{eq:gammad0}
    \Gamma_0(\mathcal{S}_1) = \{\mathcal{E}_*^u, u \in \mathcal{U}_*\}, 
\end{equation}
And $\Gamma_m(\mathcal{S}_1)$ contains all sets in $\Gamma_0(\mathcal{S}_1)$ that are maximal under inclusion.
Then we have the following graph-based characterization which is proved in Appendix~{H} in~\cite{Yuan2023}. 

\begin{theorem}
\label{thm:rdfmin}
\begin{equation*}
R(D_{min}) = \min_{\substack{p(u|s_1):S_1 \in U \in \Gamma_m(\mathcal{S}_1)}}I(U;S_1)-I(U;S_2).
\end{equation*}
\end{theorem}


\begin{remark}
Let $D = 0$ and $d_{\epsilon}(z,\hat{z}) = \mathds{1}\{ d(z,\hat{z}) > \epsilon \}$ for any $\epsilon \geq 0$, where $\mathds{1}$ denotes the indicator function. 
It is easy to check that \Cref{thm:rdfmin} is valid for both discrete and continuous alphabet~$\hat{\mathcal{Z}}$. 
Then the main result of~\cite[Theorem 3]{Basu2020} can be obtained by applying  \Cref{thm:rdfmin} to the distortion measure $d_{\epsilon}$.
\end{remark}

\begin{remark}
Assume $\mathcal{Z} = \hat{\mathcal{Z}}$ and $d$ satisfies 
\begin{equation}
    d(z,\hat{z})= 0 \text{~~iff~~} z=\hat{z}.
\end{equation}
Then the subsets in $\Gamma_m(\mathcal{S}_1)$ reduce to maximal independent sets of the characteristic graph in \cite{Orlitsky2001} and \Cref{thm:rdfmin} reduces to Theorem~2 therein.
\end{remark}
\vspace{-0.3cm}


\subsection{Simplifying Graph Characterizations in~\eqref{eq:ccf}}
\label{sec:ccf}


\subsubsection{The Case that $S_1$ and $S_2$ have a G\'{a}cs-K\"{o}rner-Witsenhausen Common Part}
\label{subsec:ccfside}
Now suppose the two-sided state information $S_1$ and $S_2$ have a G\'{a}cs-K\"{o}rner-Witsenhausen common part (cf.~\Cref{def:GKW}), denoted by $S_0 \in \{1,...,K\}$. Let $S_0 = g_1(S_1) = g_2(S_2)$ and $\mathcal{S}_{1k} = g_1^{-1}(k)$, $k = 1,...,K$.
Then similar characterizations as in~\Cref{thm:rdfci} can be obtained.

\begin{theorem}\label{thm:ccfci}
If $S_1$ and $S_2$ have a G\'{a}cs-K\"{o}rner-Witsenhausen common part, then 
\begin{equation}\label{eq:ccfci}
C(B) = -\min_{\substack{p(u|s_1):S_1 \in \mathcal{S}_1,  U \in \mathcal{U}'_{S_1}, 
\\  \mathbb{E}[b(S_1,S_2,X_{S_1})] \leq B}}I(U;S_1) - I(U;Y,S_2),
\end{equation}
where $\mathcal{U}'_{S_1} = \mathcal{X}^{g_1^{-1}(g_1(S_1))}$.
\end{theorem}

\begin{corollary}\label{cor:ccfci}
Let $S_1 = (S_0,S_1')$ and $S_2 = (S_0,S_2')$. Then we have
\begin{equation}\label{eq:ccfcisim}
C(B) = -\min_{\substack{p(u|s_1): U = (X_{s_1'})_{s_1' \in \mathcal{S}_1'},\\  \mathbb{E}[b(S_1,S_2,X_{S_1})] \leq B}}I(U;S_1)-I(U;Y,S_2).
\end{equation}
\end{corollary}

\subsubsection{Minimum Cost Case}
Let $B = B_{min}$.
For any $u = (x_{s_1})_{s_1 \in \mathcal{S}_1}$, by~\eqref{eq:assccfd} we have 
\begin{equation*}
    \tilde{l}(s_1,u) 
=\sum_{s_2}p(s_2|s_1)b(x_{s_1},s_1,s_2),
\end{equation*}
which implies 
\begin{equation}
\label{eq:lformula}
\min_{u \in \mathcal{U}}\left\{\tilde{l}(s_1,u)\right\} = \min_{x \in \mathcal{X}}\sum_{s_2}p(s_2|s_1)b(x,s_1,s_2).
\end{equation}
For any $s_1 \in \mathcal{S}_1$, define
\begin{equation}
\label{eq:defXs1}
    \mathcal{X}_{s_1} = \mathop{\arg\min}_{x \in \mathcal{X}} \sum_{s_2}p(s_2|s_1)b(x,s_1,s_2),
\end{equation}
which is the set of $x$ that achieves the optimal value.

Then for $u = (x_{s_1})_{s_1 \in \mathcal{S}_1}$, we have
\begin{equation}
\label{eq:Bmincon}
    \tilde{l}(s_1,u) = \min_{u' \in \mathcal{U}}\left\{\tilde{l}(s_1,u')\right\} \text{~~iff~~} x_{s_1} \in \mathcal{X}_{s_1}.
\end{equation}
It is intuitive that $|\mathcal{X}_{s_1}|$ is usually much smaller than $|\mathcal{X}|$.
In view of this, we can further simplify the graph characterization for $B = B_{min}$ in the following theorem. It is different from~\Cref{thm:rdfmin}, and the proof is given in Appendix~{I} in~\cite{Yuan2023}. 

\begin{theorem}
\label{thm:ccfmin}
$$C(B_{min}) = -\min_{\substack{p(u|s_1): \\ U \in \prod_{s_1 \in \mathcal{S}_1} \mathcal{X}_{s_1}}}I(U;S_1)-I(U;Y,S_2).$$
\end{theorem}

The above optimization problem is a special case of~\eqref{p0} and the number of decision variables is reduced significantly since $|\mathcal{S}_1| \cdot \prod_{s_1 \in \mathcal{S}_1}|\mathcal{X}_{s_1}| \ll |\mathcal{S}_1| \cdot |\mathcal{X}|^{|\mathcal{S}_1|}$ in general.

\begin{remark}
\label{rem:limitedpower}
    Let $B = 0$ and $b_{c}(x,s_1,s_2) = \mathds{1}\{ b(x,s_1,s_2) > c \}$ for any $c \geq 0$. Then~\Cref{thm:ccfmin} characterizes the capacity for a channel with limited power.
\end{remark}

\section{Alternating Minimization Algorithms}
\label{sec:alg}

In this section, we aim at solving the problem \eqref{p0} to obtain~$T(L)$. In light of \Cref{thm:classify0}, it suffices to confine to $L \in [L_{min},L_{Max}]$. 

\subsection{The Flexible Alternating Minimization Algorithm}
\label{subsec:givenL}

We only need to solve the equivalent form~\eqref{p2} discussed in \Cref{sec:property}.
To derive the algorithm, the Lagrange multiplier $s$ is introduced for the linear loss constraint~\eqref{p2c}.
First fix $L \in (L_{min},L_{Max})$. 
Let  $s^*$ be the corresponding Lagrange multiplier satisfying the optimality condition in the problem~\eqref{p2} with loss constraint $L$.
Since $s^*$ is unknown, the BA type approach  fixes $s$ to be some  $s'$.
But $s' \neq s^*$, hence the rate-loss function for a given $L$ cannot be computed directly following the BA approach. 
Motivated by \cite{Wu2022,Hayashi2023,Chen2023}, we overcome the weakness by updating $s$ properly.

To design the algorithm, we first construct the Lagrange function as follows.
\begin{definition}
\label{def:Lagrange}
For a fixed $s>0$, the Lagrange function is defined as
\begin{equation}
\begin{aligned}
F_s(\bm{q},\bm{r}) &\triangleq GD_{\mathcal{E}}(\bm{q}||\bm{r})+ s \cdot Loss(\bm{q}).
\end{aligned}
\end{equation}
\end{definition}

Compared with traditional rate optimization problems (such as the classical rate-distortion problem), our graph-based problem~\eqref{p2} consists of graph constraints in~\eqref{p2b}.
Hence the partial minimization process for $F_s(\bm{q},\bm{r})$ depends on the edge set $\mathcal{E}$, in contrast with similar alternating steps for those traditional problems without graph constraints.
The process is shown as follows {and proved in Appendix~J in~\cite{Yuan2023}}. 

\begin{definition}
\label{def:parmin}
For any $(v,u) \in \mathcal{E}$ and $w \in \mathcal{W}$, define the partial minimization process for $\bm{q}$,
\begin{equation}
\begin{aligned}
&\bm{q}_s^*(\bm{r})(u|v) \\
&\triangleq\frac{ e^{-s \tilde{l}(v,u)}\prod_{w'} r(u|w')^{p(w'|u,v)}}{\sum_{u' \in \mathcal{E}_v} e^{-s \tilde{l}(v,u')}\prod_{w'} r(u'|w')^{p(w'|u',v)}},\label{eq:alterq}
\end{aligned}
\end{equation}
\newline
and for $\bm{r}$,

\begin{align}
&\bm{r}^*(\bm{q})(u|w)\triangleq \frac{\sum_{v\in \mathcal{E}^u} p(v)q(u|v)p(w|u,v)}{\sum_{(v,u)\in \mathcal{E}} p(v)q(u|v)p(w|u,v)}.\label{eq:alterr}
\end{align}
Note that~\eqref{eq:alterr} does not depend on $s$. 
\end{definition}

\textbf{BA Type Algorithm: }
Fixing $s^{(n)}$ to be a positive constant~$s'$, then we can obtain a BA type algorithm with the iteration step 
\begin{align*}
    \bm{r}^{(n)} &= \bm{r}^*(\bm{q}^{(n)}),
    \\
    \bm{q}^{(n+1)} &= \bm{q}_{s'}^*(\bm{r}^{(n)}).
\end{align*}

\textbf{Our Flexible Alternating Minimization Algorithm: }
We also follow the alternating minimization approach, but in order for the algorithm to output an optimal solution for~\eqref{p2}, we update $s$ to descend $F_{s^*}(\bm{q},\bm{r})$. To be precise, the alternating step is 
\begin{equation}
\label{eq:generalalter}
\begin{aligned}
    &\bm{r}^{(n)} = \bm{r}^*(\bm{q}^{(n)}),
    \\
    &\text{choose } s^{(n+1)},
    \\
    &\bm{q}^{(n+1)} = \bm{q}_{s^{(n+1)}}^*(\bm{r}^{(n)}).
\end{aligned}
\end{equation}

To choose the suitable $s^{(n)}$, first define 
\begin{equation}
    G_{\bm{r}}(s) \triangleq Loss(\bm{q}_s^*(\bm{r})),
\end{equation}
which can be explicitly written as 
\begin{equation*}
    G_{\bm{r}}(s) = \sum_{v}p(v) \frac{ \sum_{u \in \mathcal{E}_v} \tilde{r}(u,v) e^{-s\tilde{l}(v,u)}\tilde{l}(v,u)}{\sum_{u \in \mathcal{E}_v} \tilde{r}(u,v)e^{-s \tilde{l}(v,u)}},
\end{equation*}
where $\tilde{r}(u,v) = \prod_{w'} (r(u|w'))^{p(w'|u,v)}$.

Let $\Theta_i(v) = \sum_{u\in \mathcal{E}_v}\tilde{r}(u,v)e^{-s\tilde{l}(v,u)}(\tilde{l}(v,u))^i,i = 0,1,2$, then by Cauchy-Schwarz inequality,
\begin{align*}
G_{\bm{r}}'(s) = \sum_{v}p(v) \frac{ (\Theta_1(v))^2-\Theta_0(v)\Theta_2(v)}{(\Theta_0(v))^2} \leq 0,
\end{align*}
which implies $G_{\bm{r}}(s)$ is non-increasing. 
Suppose that $\tilde{r}(u,v)>0, \forall v,u$, then 
\begin{align*}
\lim_{s \to \infty}G_{\bm{r}}(s) = \sum_v p(v) \min_{u\in \mathcal{E}_v} \big\{\tilde{l}(v,u)\big\} = L_{min},
\\
\lim_{s \to - \infty}G_{\bm{r}}(s) = \sum_v p(v) \max_{u\in \mathcal{E}_v} \big\{\tilde{l}(v,u)\big\}= L_{Max}.
\end{align*}
Therefore, the equation
\begin{equation}\label{eq:G}
    G_{\bm{r}}(s) = L,
\end{equation}
has a root for $L \in (L_{min},L_{Max})$.


If we know that $L \in (L_{min},L_{max}) \subseteq (L_{min},L_{Max}]$, we can solve the equation~\eqref{eq:G} by Newton's method.
However, since the computation of $L_{max}$ is invalid in general, we develop a practical update strategy of $s^{(n)}$ to handle~\eqref{p2c} 
for any $L \in (L_{min},L_{Max})$.

\begin{definition}
\label{strategy1}
Evaluate $G_{\bm{r}^{(n-1)}}(0)$ and define $s^{(n)}$ as follows.
\begin{enumerate}[i)]
    \item If $G_{\bm{r}^{(n-1)}}(0) \leq L$, then let $s^{(n)} = 0$.
    \item If $G_{\bm{r}^{(n-1)}}(0) > L$, then the root of $G_{\bm{r}^{(n-1)}}(s) = L$ is positive, we solve it by Newton's method and assign the solution to $s^{(n)}$.
\end{enumerate}
\end{definition}
Note that we can always obtain a nonnegative $s^{(n)}$ from~\Cref{strategy1}, i.e., the strategy is structure-preserving.

The numerical computation of $GD_{\mathcal{E}}(\bm{q}^{(n+1)}||\bm{r}^{(n)})$ through the definition in~\eqref{eq:def-GD} is not stable. In light of~\eqref{eq:alterq}, it is easy to verify that $GD_{\mathcal{E}}(\bm{q}^{(n+1)}||\bm{r}^{(n)})$ can be approximated through 
\begin{equation}
\label{eq:returnvalue}
\begin{aligned}
     &GD_{\mathcal{E}}(\bm{q}^{(n+1)}||\bm{r}^{(n)}) \approx -s^{(n)} L
     \\
     &-\!\sum_v p(v)\! \log\!\sum_{u' \in \mathcal{E}_v}\! e^{-s^{(n)} \tilde{l}(v,u')}\!\prod_{w'} r^{(n)}(u'|w')^{p(w'|u',v)}.
\end{aligned}
\end{equation}

Next we consider the boundary cases with $L = L_{min}$ or $L_{Max}$. For $L = L_{Max}$, by the definition~\eqref{eq:L_Max}, the loss constraint in~\eqref{p2c} is satisfied automatically. For $L= L_{min}$, it suffices to solve the problem with the edge set $\mathcal{E}$ replaced by~$\mathcal{E}_*$, where $(v,u) \in \mathcal{E}_*$ only if
\begin{equation}
    \tilde{l}(v,u) = \min_{u' \in \mathcal{E}_v}\left\{\tilde{l}(v,u')\right\}.
\end{equation}
Thus for both cases, there is no need to introduce the Lagrange multiplier $s$. Then the partial minimization process for $\bm{q}$ is replaced by 
\begin{equation}
\label{eq:alterq0}
\bm{q}_{0}^*(\bm{r})(u|v)=\frac{ \prod_{w'} (r(u|w'))^{p(w'|u,v)}}{\sum_{u'\in \mathcal{E}_v} \prod_{w'} (r(u'|w'))^{p(w'|u',v)}},
\end{equation}
and $\bm{r}^*(\bm{q})$ remains the same as~\eqref{eq:alterr}.

Then the Flexible Alternating Minimization Algorithm to solve~\eqref{p2} is summarized in \Cref{alg3}.

\subsection{Analysis of the Algorithm and Comparisons with Previous Methods}
\label{subsec:analysis}
The following theorem shows the convergency of~\Cref{alg3}.

\begin{figure}[!t]
\removelatexerror
\begin{algorithm}[H]
\renewcommand{\algorithmicrequire}{\textbf{Input:}}
\renewcommand{\algorithmicensure}{\textbf{Output:}}
\renewcommand{\thealgorithm}{1}
\caption{Flexible Alternating Minimization Algorithm}
\label{alg3}
\begin{algorithmic}[1]
    \REQUIRE Loss matrix $\tilde{l}(v,u)$, distributions $p(v)$,  $p(w|u,v)$, maximum iteration number $max\_iter$, loss constraint $L \in [L_{min},L_{Max}]$.
    \ENSURE An optimal solution and  the optimal value for \eqref{p2}.
    \IF{$L_{min}< L < L_{Max}$}
    \STATE Initialize  $q^{(1)}(u|v) = \frac{\mathds{1}(u \in \mathcal{E}_v)}{|\mathcal{E}_v|},\forall (v,u) \in \mathcal{E}$.
    \FOR{ $n = 1 : max\_iter$}
    \STATE $\bm{r}^{(n)} = \bm{r}^*(\bm{q}^{(n)})$ by~\eqref{eq:alterr}.
    \STATE Solve $s^{(n+1)}$ by \Cref{strategy1}.
    \STATE $\bm{q}^{(n+1)} = \bm{q}_{s^{(n+1)}}^*(\bm{r}^{(n)})$ by~\eqref{eq:alterq}.
    \ENDFOR

    \ELSE
    \IF{$L = L_{min}$}
    \STATE Override $\mathcal{E}$ by $\mathcal{E}_*$.
    \ENDIF
    \STATE Initialize $q^{(1)}(u|v) = \frac{\mathds{1}(u \in \mathcal{E}_v)}{|\mathcal{E}_v|},\forall (v,u) \in \mathcal{E}$.
    \FOR{ $n = 1 : max\_iter$}
    \STATE $\bm{r}^{(n)} = \bm{r}^*(\bm{q}^{(n)})$ by~\eqref{eq:alterr}.
    \STATE $\bm{q}^{(n+1)} = \bm{q}_{0}^*(\bm{r}^{(n)})$ by~\eqref{eq:alterq0}.
    \ENDFOR 
    
    \ENDIF
    \RETURN $(\bm{q}^{(n+1)},\bm{r}^{(n)})$ and $GD_{\mathcal{E}}(\bm{q}^{(n+1)}||\bm{r}^{(n)})$ (cf.~\eqref{eq:returnvalue}).
\end{algorithmic}
\end{algorithm}
\end{figure}


\begin{theorem}
\label{thm:convmain}
The solutions $(\bm{q}^{(n+1)},\bm{r}^{(n)})$ generated by \Cref{alg3} converge to an optimal solution $(\bm{q}^{0},\bm{r}^{0})$ and \begin{equation}
\label{eq:flexiblerate}
\begin{aligned}
    \min_{1 \leq k \leq n}F_{s^*}(\bm{q}^{(k+1)},\bm{r}^{(k)})-\min_{\bm{q},\bm{r}}F_{s^*}(\bm{q},\bm{r}) = \\ O\left(\frac{\log |\mathcal{U}|}{n}\right).
\end{aligned}
\end{equation}
Furthermore, for $L = L_{min}$ and $L = L_{Max}$,
\begin{equation}
GD_{\mathcal{E}}(\bm{q}^{(n+1)}||\bm{r}^{(n)})-GD_{\mathcal{E}}(\bm{q}^0||\bm{r}^0) = O\left(\frac{\log|\mathcal{U}|}{n}\right).
\end{equation}
\end{theorem}

{
\begin{IEEEproof}[Sketch of the Proof]
\Cref{thm:convmain} mainly relies on the following estimate of the optimality gap by~\Cref{alg3},
\begin{equation*}
\begin{aligned}
&\sum_{k = n+1}^{m}F_{s^*}(\bm{q}^{(k)},\bm{r}^{(k-1)})-F_{s^*}(\bm{q}^{0},\bm{r}^{0})
\\
&\leq GD_{\mathcal{E}}(\bm{q}^0||\bm{q}^{(n)})-GD_{\mathcal{E}}(\bm{q}^0||\bm{q}^{(m)}) \leq \log |\mathcal{U}|
\end{aligned}
\end{equation*}
for any $m \geq n \geq 1$.
It can be established mainly thanks to the flexible update strategy in~\Cref{strategy1}.
Then the estimate for algorithms in~\cite{Blahut1972,Arimoto1972,Chen2023} focusing on problems without side information can be generalized to~\Cref{alg3} for the unified problem~\eqref{p0}, which also subsumes problems with side information and graph constraints~\eqref{p0b}. 
Detailed proof can be found in Appendix~K of the complete version~\cite{Yuan2023}. 
\end{IEEEproof}
}

The computational complexity of each iteration in these algorithms is proportional to $|\mathcal{U}|\cdot|\mathcal{V}|\cdot|\mathcal{W}|$. 
By~\Cref{thm:convmain} $O(\frac{\log|\mathcal{U}|}{\epsilon})$ iterations are sufficient to compute the optimal value to an error $\epsilon$.
Consequently, to achieve an accuracy $\epsilon$, the total computation cost by~\Cref{alg3} is bounded by $O(\frac{|\mathcal{U}||\mathcal{V}||\mathcal{W}|\cdot \log|\mathcal{U}|}{\epsilon})$.

Considering $\mathcal{V}$ and $\mathcal{W}$ are always fixed by the problem, the only available approach for reducing the computational complexity is to reduce $|\mathcal{U}|$.
This can be achieved by exploiting the simplified graph characterizations in~\Cref{sec:rdfccf} and introducing deflation techniques, detailed in~\Cref{subsec:alggraph,subsec:deflation} respectively.

\begin{remark}
Similar to our Flexible Alternating Minimization Algorithm, the BA type algorithm in~\Cref{subsec:givenL} can be shown to be convergent as well, see~Lemma~13 in Appendix~{K} in~\cite{Yuan2023} for details.
\end{remark}

\begin{remark}
\label{rem:rdf}
Consider the rate-distortion problem~\eqref{eq:rdf}, which is a special case of our problem~\eqref{p0}. 
In this case, our algorithm can be simplified as follows.
\begin{enumerate}
    \item Recall that the explicit value of~$L_{max}$ cannot be directly computed in the general problem~\eqref{p0}. While for~\eqref{eq:rdf}, we can verify that $D_{max} = \min_{u \in \mathcal{U}} \left\{\sum_{s_1} p(s_1) \tilde{l}(s_1,u)\right\}$, and $R(D) = 0$ for $D \geq D_{max}$. Since we can calculate~$D_{max}$, it suffices to confine to $D \in (D_{min},D_{max})$. Hence in~\Cref{strategy1}, we can simply obtain the root of $G_{\bm{r}^{(n-1)}}(s) = L$ by Newton's method and assign the value to~$s^{(n)}$. 
\item By~\eqref{eq:assrdfd} the partial minimization process~\eqref{eq:alterr} for $\bm{r}$ is reduced to
\begin{equation*}
\bm{r}^*(\bm{q})(u|s_2)= \sum_{s_1\in \mathcal{E}^u} q(u|s_1)p(s_1|s_2).
\end{equation*}
\end{enumerate}
\end{remark}

\begin{remark}
Further specializing to the classical rate-distortion problem, then the algorithm in~\Cref{rem:rdf} reduces to the algorithm in~\cite{Hayashi2023,Chen2023}. 
Note that inequality constraints cannot be reduced to equality constraints in general (because $L_{max}$ is not available) especially for the capacity-cost problem in~\eqref{eq:ccf}, hence the algorithm in~\Cref{rem:rdf} cannot be applied to the general problem~\eqref{p0}.
Also, methods in~\cite{Hayashi2023,Chen2023} for handling equality constraints are not sufficient to solve our problem~\eqref{p0}.
\end{remark}

\begin{remark}
For the lossy computing problem with decoder-side information with zero distortion, \Cref{alg3} for $L = L_{min}$ can be specialized to the BA type algorithm computing the conditional graph entropy in~\cite{Harangi2022}.
\end{remark}

\begin{remark}
Compared with the methods in~\cite{Cheng2005,Dupuis2004,Willems1983} designed for specific rate optimization problems with side information, our method has many advantages.

1) Our methods apply to a much wider class of problems. 
Previous works~\cite{Cheng2005,Dupuis2004,Willems1983} designed algorithms for a special case of the rate-distortion problem in~\Cref{eg:rdf} with $f(x,y) = x$ and the capacity-cost problem in~\Cref{eg:ccf}, respectively. 
Algorithms for the rate-distortion problem of computing a general function in~\Cref{eg:rdf} or with causal side information in~\Cref{eg:crdf} have not been designed.
Through introducing and computing a unified graph-based rate optimization problem~\eqref{p0}, all these problems can be solved by~\Cref{alg3}.

2) Our algorithm is more flexible than previous methods in~\cite{Cheng2005,Dupuis2004,Willems1983} based on the BA 
approach. 
As we have noted in~\Cref{subsec:givenL}, a BA type algorithm cannot compute the problem~\eqref{p2} directly, because the Lagrange multiplier
$s$ is a real input of the
algorithm, but the loss criterion $L$ is not. 
As a consequence,
to compute the problem~\eqref{p2} for the fixed $L$, one has to add an outer iterative procedure to search
for the Lagrange multiplier, invoking the BA type algorithm
as a subroutine. This incurs high computational complexity.
In contrast, our algorithm can output the optimal value and an optimal solution to~\eqref{p2} directly.

3) For the problem that can be solved by the method in~\cite{Cheng2005}, our computational complexity is much lower than~\cite{Cheng2005}. 
Take the capacity-cost problem~\eqref{eq:ccf} as an example, we have $|\mathcal{U}| = |\mathcal{X}|^{|\mathcal{S}_1|}$ and the total complexity is 
$O(\frac{|\mathcal{X}|^{|\mathcal{S}_1|}|\mathcal{S}_1|^2|\mathcal{S}_2||\mathcal{Y}|\cdot \log|\mathcal{X}|}{\epsilon})$ for our method to achieve an accuracy $\epsilon$. 
The method by~\cite{Cheng2005} started from a direct characterization
\begin{equation*}
    C(B) = - \min I(U';S_1)-I(U';Y,S_2).
\end{equation*}
where the minimum is taken over all functions $g$ and transition probabilities $p(u'|s_1)$ such that $ \mathbb{E}[b(g(U',S_1),S_1,S_2)] \leq B$.
For each fixed input function $g$, $|\mathcal{U}'| = \min\{|\mathcal{X}|\cdot |\mathcal{S}_1|+1, |\mathcal{Y}|+|\mathcal{S}_1|\}$ and a BA type iteration was used to optimize $p(u'|s_1)$.
But the method needs to traverse all $g$, resulting in an additional factor $|\mathcal{X}|^{|\mathcal{S}_1|\cdot|\mathcal{U}'|}$ in the complexity. 
Even ignoring additional factors for searching the multiplier $s^*$, The total complexity has become $O(\frac{|\mathcal{X}|^{|\mathcal{S}_1|\cdot|\mathcal{U}'|}|\mathcal{U}'||\mathcal{S}_1||\mathcal{S}_2||\mathcal{Y}|\cdot \log|\mathcal{U}'|}{\epsilon})$, which is much higher than our method.

4) Furthermore, our algorithm can be more efficient with the help of graph characterizations and deflation techniques. Both of them can be applied to BA type algorithm as well, but they were not exploited by previous works~{\cite{Cheng2005,Dupuis2004,Willems1983,Harangi2022}}.
\end{remark}

\subsection{Exploiting the Simplified Graph Characterizations}
\label{subsec:alggraph}

The specific structures of graph characterizations in~\Cref{sec:rdf,sec:ccf} can be useful.
The problem is simplified by
reducing the alphabet $|\mathcal{U}|$, and hence the number of decision variables $\bm{p}_{U|V}$.
By inputting simplified versions of~\eqref{p0} in~\Cref{alg3}, the computational complexity is greatly reduced. 

It suffices to show that these simplified versions of~\eqref{p0} can be efficiently computed.
First, consider the characterizations in~\Cref{thm:rdfci,thm:ccfci} in cases that two-sided information $S_1$ and $S_2$ has a nontrivial G\'{a}cs-K\"{o}rner-Witsenhausen common part. 
To compute the GKW common part, 
first construct a bipartite graph $\mathscr{G}[S_1,S_2,\mathscr{E}]$, where $(s_1,s_2) \in \mathscr{E}$ if $p(s_1,s_2)>0$. 
Then by~\Cref{def:GKW}, determining the GKW common part is equivalent to finding all the connected components of the bipartite graph $\mathscr{G}[S_1,S_2,\mathscr{E}]$.
The latter task can be effectively completed by a graph traversal algorithm with  complexity no larger than $O(|\mathcal{S}_1||\mathcal{S}_2|)$. 
Then the characterizations~\eqref{eq:rdfci} and~\eqref{eq:ccfci} is immediately obtained.

For the minimum loss case, the graph characterizations in~\Cref{thm:rdfmin,thm:ccfmin} can be obtained by directly performing the graph operations discussed in~\Cref{sec:rdf,sec:ccf}, the complexity is again no larger than $O(|\mathcal{S}_1||\mathcal{S}_2|)$.

To see the reduction of complexity by graph characterizations, take the computation of the rate-distortion function as an example. If the two-sided information has a G\'{a}cs-K\"{o}rner-Witsenhausen common part for instance $S_1 = (S_0,S_1')$ and $S_2 = (S_0,S_2')$, then  $|\mathcal{U}|$ can be reduced from $ |\hat{\mathcal{Z}}|^{|\mathcal{S}_2|}$ to $|\hat{\mathcal{Z}}|^{|\mathcal{S}'_2|} $. 
The induced block structure of the tensor $p(w|u,v)$ can be exploited to further reduce the cost of each iteration by a factor~$|\mathcal{S}_0|$.
The total complexity is reduced from $O(\frac{|\hat{\mathcal{Z}}|^{|\mathcal{S}_2|}|\mathcal{S}_1||\mathcal{S}_2|^2\cdot \log|\hat{\mathcal{Z}}|}{\epsilon})$ to $O(\frac{|\hat{\mathcal{Z}}|^{|\mathcal{S}_2'|}|\mathcal{S}_1||\mathcal{S}'_2|^2\cdot \log|\hat{\mathcal{Z}}|}{\epsilon})$, if the accuracy $\epsilon$ is achieved.
We see that a factor $|\mathcal{S}_0|$ is eliminated in the exponential, which is a significant acceleration. 

\subsection{Deflation Techniques}
\label{subsec:deflation}
Graph characterizations in \Cref{sec:rdf,sec:ccf} have reduced the number of decision variables. However, the number is commonly still much larger than $|\mathcal{V}|$ and $|\mathcal{W}|$, e.g., $|\mathcal{U}| = |\hat{\mathcal{Z}}|^{|\mathcal{S}_2|}$ and $|\mathcal{X}|^{|\mathcal{S}_1|}$ for~\eqref{eq:rdf} and~\eqref{eq:ccf}.  This results in a very large computational cost for each iteration. We can reduce the cost due to the existence of sparse solutions in the following sense. {The proof is by the support lemma, which can be found in Appendix~L in~\cite{Yuan2023}.}

\begin{lemma}
\label{lem:sparsity}
    For the problem \eqref{p0}, 
    there exists some optimal solution $p(u|v)$ such that there are at most $|\mathcal{V}|+|\mathcal{W}|$ of $u$ with $p(u)>0$.
\end{lemma}

%


An optimal solution $p(u|v)$ of \eqref{p0} with a sparse support of $U$ can be obtained as follows. It is intuitive that $u$ with a smaller cost should have a larger probability. 
Then we can regard our algorithm as a feature enhancement process with $u$ being the feature. The algorithm begins with a fixed $p(u|v)$ for each $v \in \mathcal{V}$, then $p(u)$ is roughly averaged over $\mathcal{U}$. During iterations, the algorithm enhances the feature $u$ that has a smaller cost by increasing $p(u)$. 
The probability of $u$ with a large cost will finally converge to $0$.

In practice, we perform deflation techniques as follows. 
\begin{definition}
\label{def:defla}
Choose a suitable $k \in \mathbb{N}$ and a small constant $\delta > 0$ at the beginning of the algorithm.
For the $n$-th iteration, assume that $\bm{q}^{(n+\frac{1}{2})}$ is computed by~\eqref{eq:alterq}.
If $n = k-1 (mod~k)$, then delete each $u \in \mathcal{U}$ with 
$$
p^{(n+\frac{1}{2})}(u)< \delta/|\mathcal{U}|,
$$
where $p^{(n+\frac{1}{2})}(u) = \sum_{v \in \mathcal{E}^u}q^{(n+\frac{1}{2})}(u|v)p(v)$.
After that, renormalize $\bm{q}^{(n+\frac{1}{2})}$ to obtain $\bm{q}^{(n+1)}$.
\end{definition}
In other words, in~\Cref{def:defla} we can update the support of $U$ periodically by deleting $u$ if $p^{(n+\frac{1}{2})}(u)$ becomes very small.
The accelerated algorithm with both simplified graph characterizations 
 and deflation techniques is summarized in~\Cref{alg5} (line~1 and 6-13).

\begin{figure}[!t]
\removelatexerror
\begin{algorithm}[H]
\renewcommand{\algorithmicrequire}{\textbf{Input:}}
\renewcommand{\algorithmicensure}{\textbf{Output:}}
\renewcommand{\thealgorithm}{2}
\caption{Flexible Alternating Minimization Algorithm with Acceleration Techniques}
\label{alg5}
\begin{algorithmic}[1]
    \REQUIRE Loss matrix $\tilde{l}(v,u)$, distributions $p(v)$, $p(w|u,v)$, maximum iteration number $max\_iter$, deflation period~$k$, deflation threshold $\delta$.
    \ENSURE Optimal value and an optimal solution.
    \STATE Simplify the graph-based problem as in~\Cref{subsec:alggraph} and update $\tilde{l}(v,u)$, $p(v)$, $p(w|u,v)$ accordingly.
    \STATE Initialize $q^{(1)}(u|v) = \frac{\mathds{1}(u \in \mathcal{E}_v)}{|\mathcal{E}_v|}, \forall (v,u) \in \mathcal{E}$.
    \FOR{ $n = 1 : max\_iter$}
    \STATE $\bm{r}^{(n)} = \bm{r}^*(\bm{q}^{(n)})$ by~\eqref{eq:alterr}.
    \STATE Solve $s^{(n+1)}$ by \Cref{strategy1} as in~\Cref{alg3}.
    \STATE $\bm{q}^{(n+\frac{1}{2})} = \bm{q}_{s^{(n+1)}}^*(\bm{r}^{(n)})$ by~\eqref{eq:alterq}.
    \IF{$n = k-1 (\mathrm{mod}\  k)$}
    \STATE $\mathcal{U} \leftarrow \{u \in \mathcal{U} |\sum_{v \in \mathcal{E}^u}q^{(n+\frac{1}{2})}(u|v)p(v)\geq \frac{\delta}{|\mathcal{U}|} \}$.
    \STATE $\mathcal{E} \leftarrow \mathcal{E} \cap \mathcal{V} \times \mathcal{U}$.
    \STATE $q^{(n+1)}(u|v) = \frac{q^{(n+\frac{1}{2})}(u|v)}{\sum_{u' \in \mathcal{E}_v}q^{(n+\frac{1}{2})}(u'|v)}, \forall (v,u) \in \mathcal{E}$.
    \ELSE 
    \STATE $\bm{q}^{(n+1)} = \bm{q}^{(n+\frac{1}{2})}$.
    \ENDIF
    \ENDFOR
    \RETURN $(\bm{q}^{(n+1)},\bm{r}^{(n)})$ and $GD_{\mathcal{E}}(\bm{q}^{(n+1)}||\bm{r}^{(n)})$ (cf.~\eqref{eq:returnvalue}).
\end{algorithmic}
\end{algorithm}
\end{figure}

Since shrinkage of the alphabet of $U$ only makes the feasible region of the problem~\eqref{p2} smaller, \Cref{alg5} with deflation techniques always outputs a feasible solution of~\eqref{p2} and an upper bound (or achievable rate) for the optimal value.
By letting $\delta \to 0$, the bound can approximate the optimal value of~\eqref{p2}. 

Note that the deflation techniques can be applied to accelerate both~\Cref{alg3} and the  BA type algorithm in~\Cref{subsec:givenL}.
The size of the support of $U$ decreases exponentially until it is comparable to $|\mathcal{V}|$ and $|\mathcal{W}|$, as verified by numerical experiments in \Cref{sec:num}. 
Then the complexity of each iteration is relatively small.
We can see from numerical experiments in~\Cref{subsec:defla} that the techniques greatly save computational time without loss of accuracy and can handle problems with larger sizes.

\section{Numerical Results and Discussions}
\label{sec:num}
This section is devoted to analyzing the performance of our algorithm by several numerical computation examples.
All the experiments are conducted on a PC with 16G RAM,
and { with} one Intel(R) Core(TM) i7-7500U CPU @2.70GHz. 

\subsection{Verification of the algorithm for Classical Problems}\label{subsec:classical}
Consider two set of classical examples. One is the rate-distortion function for the online card game in~\cite{Yuan2022}, and the other is the capacity for memory with stuck-at faults (Example~7.3 in~\cite{EIGamal2011}).
The analytical results can be found in \cite{Yuan2022} and \cite{EIGamal2011}, respectively.

For the first one, let $\mathcal{X} = \mathcal{Y} = \{1,2,3\}$, $p(i,j) = \frac{1}{6} \cdot \mathds{1}{(i \neq j)},\ i,j = 1,2,3$ and $f(x,y) = \mathds{1}\{x>y\}$. Also, $\hat{\mathcal{Z}} = \mathcal{Z} = \{0,1\}$ and the distortion measure $d$ is set to be the Hamming distortion. We consider three cases in \Cref{tab:ITWeg}.

\begin{table}[!t]
\caption{Rate-Distortion Functions for the Online Card Game}
\label{tab:ITWeg}
\centering
\begin{tabular}{|c|c|c|c|}
\hline
 $R(D)$ & $S_1$& $S_2$ & Analytical results\\
\hline
$R_1(D)$ & $X$ &$Y$&$\frac{2}{3}\mathds{1}_{\{D \leq \frac{1}{6}\}}(H(\frac{1+6D}{4})-H(3D))$\\
\hline
$R_2(D)$ &$(X,Y)$&$\emptyset$&$\mathds{1}_{\{D \leq \frac{1}{2}\}}(1-H(D))$\\
\hline
$R_3(D)$ & $(X,Y)$&$Y$&$\frac{1}{3}\mathds{1}_{\{D \leq \frac{1}{6}\}}(1-H(3D))$\\
\hline
\end{tabular}
\end{table}

For the second one, $\mathcal{X} = \mathcal{Y} = \{0,1\}$ and $S$ is the channel state with $\mathcal{S} = \{1,2,3\}$. For $S=1$ ($S=2$), the output $Y$ is always $0$ ($1$), independent of the input $X$. For $S=3$, there is no fault and $Y=X$. The probabilities of these states are $p/2$, $p/2$ and $1-p$, respectively. The capacity parameterized by the error probability $p$ for different cases is given in \Cref{tab:memoryeg}.

\begin{table}[!t]
\caption{The Capacity for memory with stuck-at faults}
\label{tab:memoryeg}
\centering
\begin{tabular}{|c|c|c|c|}
\hline
Capacity & $S_1$& $S_2$ & Analytical results\\
\hline
$C_1(p)$ & $\emptyset$ &$\emptyset$&$1-H(\frac{p}{2})$\\
\hline
$C_2(p)$ &$S$&$\emptyset$&$1-p$\\
\hline
$C_3(p)$ & $\emptyset$&$S$&$1-p$\\
\hline
$C_4(p)$ & $S$&$S$&$1-p$\\
\hline
\end{tabular}
\end{table}

\begin{figure}[!t]
    \centering
    \includegraphics[scale = 0.3]{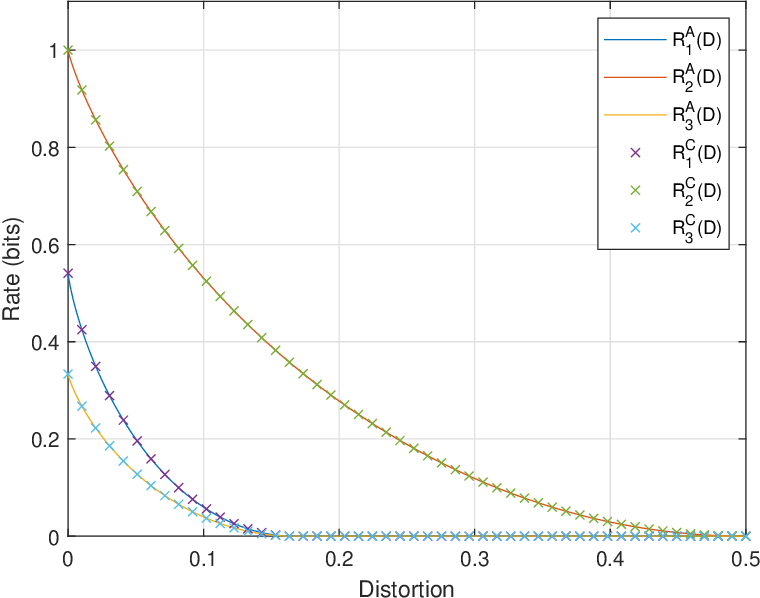}
    \includegraphics[scale = 0.3]{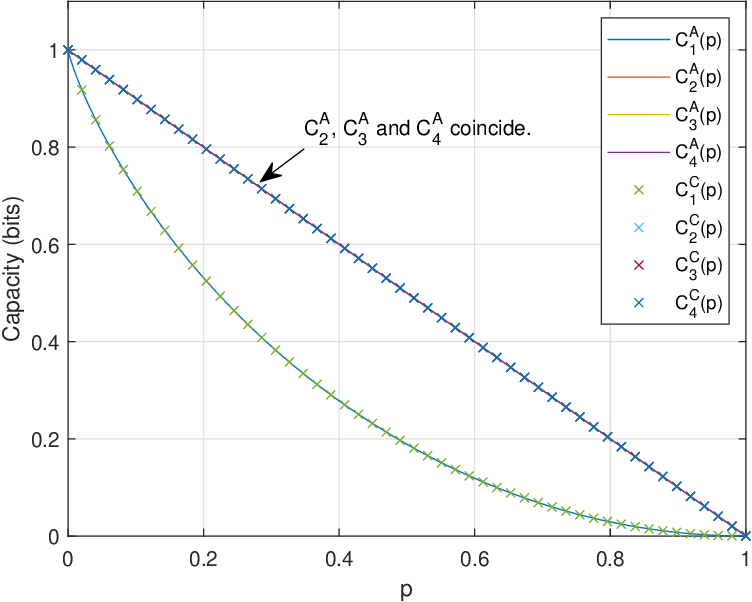}
    \caption{Analytical (superscript $A$) and Numerical results (superscript $C$) for the first (upper) and second (lower) examples in \Cref{subsec:classical}. In each case, we compute the optimal rate (capacity) with $150$ iterations for each point.}
    \label{fig:anyvsnum}
\end{figure}

In Fig.~\ref{fig:anyvsnum}, we plot the analytical curves  of the rate-distortion function and the capacity as well as points computed by~\Cref{alg3}. We observe that all the points exactly lie on the analytical curve, which shows the accuracy of our algorithm. Considering the number of iterations for each point is relatively limited ($150$ steps), the efficiency of our algorithm is also illustrated.

\subsection{Applications to Problems without Analytical Solutions}
\label{subsec:without}
Consider the rate-distortion and capacity-cost functions for some complex scenarios.  In these cases, analytical solutions have not been found and our algorithm plays an important role in numerical solutions. 

\subsubsection{Rate-distortion functions for two lossy computing problems}
\label{subsubsec:rdf}
Let $\mathcal{S}_1 =\{1,2,...,6\}$, $\mathcal{S}_2 = \{1,2,3,4\}$ and $p(i,j) = \frac{1}{24},\forall i,j $.   
We consider a common sum function $f(s_1,s_2) = s_1+s_2$ as the first example and a general nonlinear function $f(s_1,s_2) = s_1 s_2-s_2+5$ as the second example.
We set $\mathcal{Z} = \hat{\mathcal{Z}} = \{ 2,3,...,10\}$ for the first one and $\mathcal{Z} = \hat{\mathcal{Z}} = \{5,6,...,25\}$ for the second one.
In both examples, the distortion measure $d$ is set to be the quadratic distortion.

\begin{figure}[!t]
    \centering
    \includegraphics[scale = 0.3]{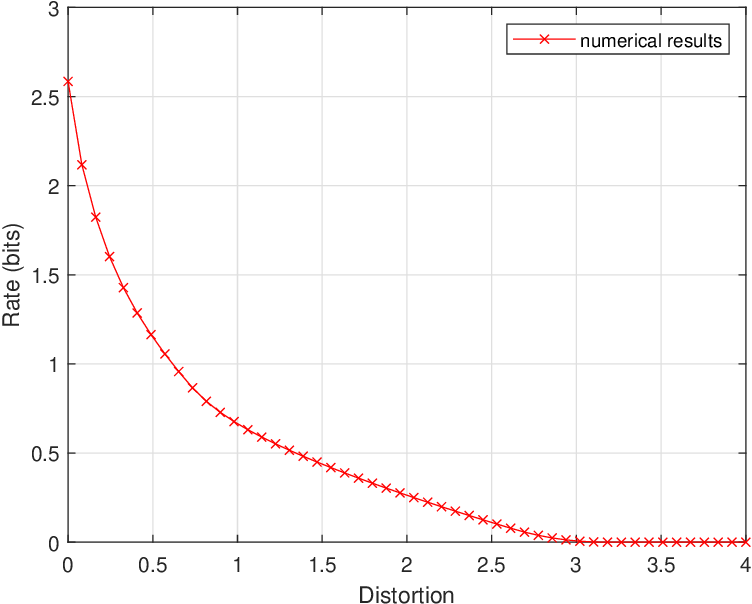}
    \includegraphics[scale = 0.3]{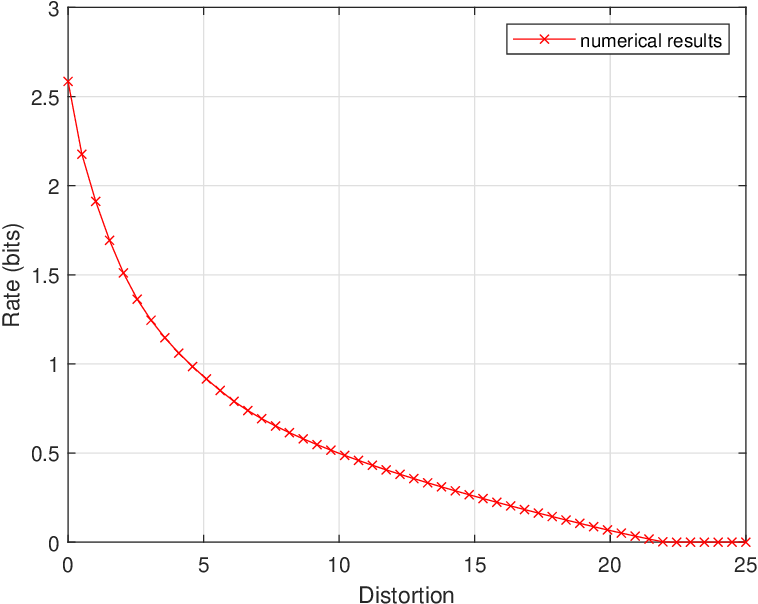}
    \caption{Numerical results for the first (upper) and second (lower) examples in \Cref{subsubsec:rdf}. In each case, we choose $50$ consecutive points from the intervals uniformly and compute the corresponding optimal rate with $1000$ iterations.  }
    \label{fig:num}
\end{figure}

We use \Cref{alg5} (specifically \Cref{alg3} with deflation techniques) to compute the rate-distortion functions for $50$ consecutive $D$ and plots the curves in Fig.~\ref{fig:num}. 

\subsubsection{Capacity-cost function for the Gaussian additive channel with quantized state information}\label{subsubsec:Gaussian}
We consider the channel 
\begin{equation}
    Y = X+S+Z,
\end{equation}
where the channel state $S \sim  \mathrm{N}(0,\frac{1}{2})$ and the noise $Z \sim \mathrm{N}(0,1)$ are independent.

A more practical situation is that $S$ is measured with a given degree of accuracy, so that a quantized version of $S$ is known by the encoder. More formally, let 
\begin{equation}
    Q_4(s) = \mathrm{sgn}(s)(1.5 \cdot \mathds{1}\{|s|>1\}+0.5\cdot\mathds{1}\{|s| \leq 1\})
\end{equation}
be the quantization function and $S_1 = Q_4(S/\sqrt{\frac{1}{2}})$ be the two-bit quantized state information.

We perform uniform quantization of $X$ and $Y$ over intervals $[-4,4]$ and $[-8,8]$, respectively. Also, $|X| = 2^b$ and $|Y| = 2^{b+1}$, which means there are $b$ bits to represent the input $X$ and $b+1$ bits to express the output~$Y$. The transition probability $p\big(y|(x_{s_1})_{s_1 \in \mathcal{S}_1},s_1\big)$ is computed by the $5$-point closed Newton-Cotes quadrature rule applied on the probability density function.

\begin{figure}[!t]
    \centering
    \includegraphics[scale = 0.575]{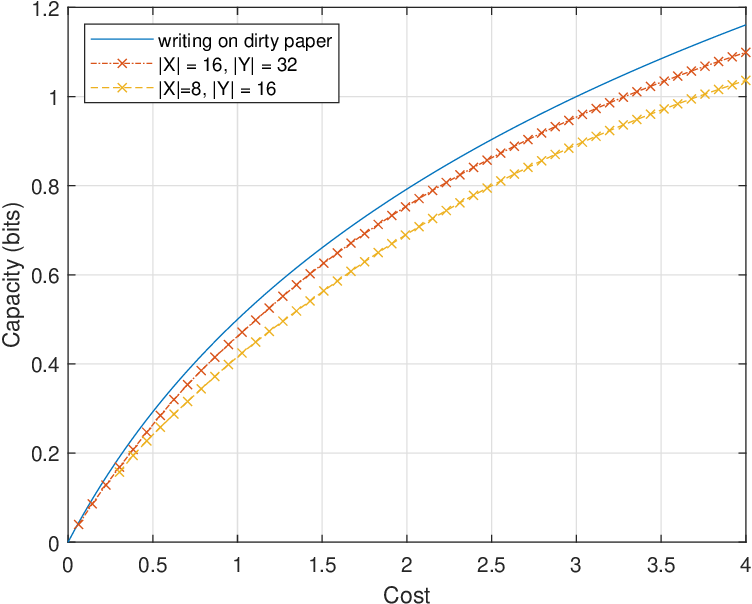}
    \caption{Numerical results for the channel problem with quantized side information in \Cref{subsubsec:Gaussian}. Capacity-cost curves for two schemes with different quantization granularity are plotted.
    In each case, $50$ consecutive points are chosen from the intervals uniformly and the corresponding capacity is computed with $1000$ iterations. The capacity-cost curve of the writing on dirty scheme is also plotted as an upper bound. }
    \label{fig:Gau}
\end{figure}

The capacity-cost function computed by \Cref{alg5}  for $b = 3,4$ is plotted in Fig. \ref{fig:Gau}. Note that if $S$ is fully known to the encoder, the capacity $\frac{1}{2} \log (1+B)$ is given by the writing on dirty paper scheme in~\cite{Costa1983}, which provides an upper bound for the capacity of the quantized version here.

\subsection{Comparisons with {Existing Algorithms}}\label{subsec:numcom}

We compare our Flexible Alternating Minimization Algorithm (FAM) with acceleration techniques (i.e.~\Cref{alg5}), with the BA type algorithm in~\Cref{subsec:givenL} designed by generalizing the methods in~\cite{Cheng2005,Willems1983,Dupuis2004} without acceleration techniques.  
The performance of two algorithms is measured by comparing their computational time over $50$ trials in Table~\ref{tab:compare}, where - means that the computational time is over $3600$ seconds and * denotes the capacity-cost function for Gaussian channel with quantized state information in~\Cref{subsubsec:Gaussian}.
The BA type algorithm cannot compute the rate directly with a given $L$, hence we perform binary search on the corresponding multiplier $s$ to ensure accuracy. It generally takes about $\log (\frac{1}{\epsilon})$ trials to search for a suitable multiplier $s$ and compute $T(L)$ to an absolute error $\epsilon$. Both algorithms are stopped until optimal values are computed to the accuracy~$\epsilon \sim 10^{-6}$. 

\begin{table}[!t] 
\centering
\caption{Comparison of the Computational Time between {the BA Type Algorithm and Our FAM Algorithm}} 
\begin{tabular}{|c|c|c|c|c|c|}
\hline
 \multicolumn{2}{|c|}{\multirow{2}*{Examples}} & \multirow{2}*{$L$}&\multicolumn{2}{c|}{Time (s)}  & Speed-up \\
\cline{4-5}
\multicolumn{2}{|c|}{}& &$t_{BA}$ & $t_{FAM}$ &ratio 
\\
\hline
\multicolumn{2}{|c|}{Sum function}  &$0.5$ &$14.12$ &$0.0843$ &$168$\\
\multicolumn{2}{|c|}{computation}  &$2.5$ &$86.46$ &$0.1108$ &$780$\\
\hline
\multicolumn{2}{|c|}{Nonlinear}  &$0.5$ &$742.4$ &$1.571$ &$473$\\
   \multicolumn{2}{|c|}{function}&$5.0$ &- &$5.515$ &-\\  \multicolumn{2}{|c|}{computation}&$20.0$ &- &$4.095$ &-\\
\hline
 &$|X| = 8$ &$1.5$ &- &$2.121$ &-\\
 &$|Y| = 16$  &$5.0$ &$271.6$ &$0.3021$ &$899$\\
\cline{2-6}
* & $|X| = 16$&$1.5$ &-&$14.05$ &-\\
& $|Y| = 32$&$5.0$ &-&$8.409$ &-\\
 \cline{2-6}
&$|X| = 32$ &$1.5$ & - &$203.1$ &-\\
&$|Y| = 64$ &$5.0$ &- &$167.5$ &-\\
\hline
\end{tabular}

\label{tab:compare} 
\end{table}

From Table~\ref{tab:compare}, we can see that our algorithm is much faster than the BA type algorithm if the problem is computed to the same order of accuracy. The advantage of our algorithm becomes more remarkable as the size of the problem gets larger.
As a result, our algorithm can compute relatively large problems (large alphabets) that the BA type algorithm fails to solve in a reasonable time.
This is clearly revealed in the computation of $R(D)$ for the nonlinear function computation problem in \Cref{subsubsec:rdf} and $C(B)$ for the channel problem in~\Cref{subsubsec:Gaussian}.

\subsection{The Effects of Acceleration Techniques}\label{subsec:defla}
We investigate the speed-up effects of~\Cref{alg5}  against \Cref{alg3} through the two examples in~\Cref{subsec:without}.
Set the deflation period $k =5$, the deflation threshold $\delta = 10^{-2}$, the number of iterations $max\_iter=1000$ for the example in \Cref{subsubsec:rdf} and $max\_iter=2000$ for the example in \Cref{subsubsec:Gaussian}. The time is averaged over $50$ experiments to eliminate the effect of noise. The computing time, speed-up ratio and loss of accuracy (the difference of the computed optimal rates with and without deflation techniques) are summarized in \Cref{tab:timeerror}, similar to~\Cref{tab:compare}.

\begin{table}[!t]
\caption{The Computing Time and Loss of Accuracy with and without Deflation Techniques for Some Examples}
\label{tab:timeerror}
\centering
\begin{tabular}{|c|c|c|c|c|c|c|}
\hline
 \multicolumn{2}{|c|}{\multirow{2}*{Examples}} & \multirow{2}*{$L$}&\multicolumn{2}{c|}{Time (s)}  & Time& Loss of  \\
\cline{4-5}
\multicolumn{2}{|c|}{}& &before & after &Ratio & Accuracy
\\
\hline
\multicolumn{2}{|c|}{Sum function}  &$0.5$ &$8.07$ &$0.09$ &$87.8$&\num{1.60e-16}\\
\multicolumn{2}{|c|}{computation}  &$2.5$ &$4.91$ &$0.13$ &$39.4$&\num{1.44e-15}\\
\hline
\multicolumn{2}{|c|}{Nonlinear}  &$0.5$ &$481$ &$1.75$ &$274$&\num{8.88e-16}\\
   \multicolumn{2}{|c|}{function}&$5.0$ &$175$ &$1.82$ &$96.3$&\num{9.97e-9}\\  \multicolumn{2}{|c|}{computation}&$20.0$ &$173$ &$3.95$ &$43.7$&\num{1.11e-8}\\
\hline
 &$|X| = 8$ &$1.5$ &$8.47$ &$0.39$ &$21.8$&\num{3.32e-10}\\
 &$|Y| = 16$  &$5.0$ &$8.40$ &$0.33$ &$25.2$&\num{2.96e-13}\\
\cline{2-7}
* & $|X| = 16$&$1.5$ &$285$ &$6.89$ &$41.4$&\num{3.16e-9}\\
& $|Y| = 32$&$5.0$ &$280$ &$6.51$ &$43.0$&\num{2.43e-9}\\
 \cline{2-7}
&$|X| = 32$ &$1.5$ & - &$175$ &-&-\\
&$|Y| = 64$ &$5.0$ &- &$162$ &-&-\\
\hline
\end{tabular}
\end{table}

We see from~\Cref{tab:timeerror} that our acceleration techniques greatly reduce the computing time at the expense of a small penalty. 
Again we find the speed-up ratio increases as the size of the problem gets larger, clearly seen from the three Gaussian cases.

The trend of time and error for different number of iterations is shown by \Cref{tab:time}. The first case is to compute $R(5.0)$ for the nonlinear function computation problem in \Cref{subsubsec:rdf} and the second case is to compute $C(1.5)$ for the channel problem with $b = 4$ in \Cref{subsubsec:Gaussian}. The value computed through sufficiently many iterations
is regarded as the true value. We take the average over $50$ experiments again. Note that we view the initialization time as the time for $0$ iteration which is inherent regardless of the algorithm.

\begin{table*}[!t]
\caption{The Time and Error of Different Number of Iterations with Deflations\label{tab:time}}
\centering
\begin{tabular}{|c|c|c|c|c|c|c|c|c|c|}
\hline
\multicolumn{2}{|c|}{Number of Iterations} &$0$&$30$&$60$& $125$ & $250$ & $500$ &$1000$&$2000$\\
\hline
Nonlinear &Time (s) &$0.184$&$1.437$ &$1.537$& $1.654$ &$1.728$&$1.761$&$1.820$ &$1.907$\\
\cline{2-10}
Function & Error &-&\num{6.42e-2} &\num{3.29e-2}& \num{1.35e-2} &\num{3.90e-3}&\num{9.22e-4}&\num{2.75e-4} &\num{6.61e-5}\\
\hline
Gaussian &Time (s) &$1.865$&$4.979$ &$5.679$& $6.184$ &$6.543$&$6.612$&$6.719$ &$6.889$\\
\cline{2-10}
channel & Error &-&\num{8.28e-2} &\num{3.53e-2}& \num{1.28e-2} &\num{3.89e-3}&\num{9.49e-4}&\num{1.84e-4} &\num{3.12e-5}\\
\hline
\end{tabular}
\end{table*}

The numerical experiments in~\Cref{tab:time}  verify our discussion in~\Cref{subsec:deflation} that the time of each iteration decreases sharply very soon. This can be seen from the table that the total time increases slower and slower and thus thousands of iterations can be applied to achieve a higher accuracy. 


\balance
\bibliographystyle{bibliography/IEEEtran}
\bibliography{bibliography/computing}

\clearpage
\nobalance
\twocolumn[{
\begin{center}
  \Huge Supplement to ``Computation of a Unified Graph-Based Rate Optimization Problem''
\end{center}
\begin{center}
  \large 
    Deheng Yuan, Tao Guo, {\it Member, IEEE},
Zhongyi Huang and Shi Jin, {\it Fellow, IEEE}%
\end{center}
\vskip 4em
}]

\setcounter{page}{1}
\appendices

We present in this supplementary
material the detailed proofs of~\Cref{lem:rdf,lem:ccf,lem:crdf,lem:con opt,lemma:T(L),lem:sparsity} and \Cref{thm:rdfci,thm:rdfmin,thm:ccfmin,thm:convmain} in the paper ``Computation of a Unified Graph-based Rate Optimization Problem''.

\section{Graph Contractions}
\label{sec:graph}

We introduce graph operations to contract the feasible region in~\eqref{lem:pgraph}, preparing for the proof of~\Cref{lem:rdf,lem:ccf} and~\Cref{thm:rdfci,thm:rdfmin,thm:ccfci,thm:ccfmin}. 

Let $\mathcal{H}' \subseteq \mathcal{H}$,  $\mathcal{F}' \subseteq \mathcal{F} \cap \mathcal{V} \times \mathcal{H}'$ and $G'[\mathcal{V},\mathcal{H}',\mathcal{F}']$ be a {\it subgraph} of $G[\mathcal{V},\mathcal{H},\mathcal{F}]$.

\begin{definition}
\label{def:feacon}
A {\it feasible contraction} from $\big(G[\mathcal{V},\mathcal{H},\mathcal{F}],\omega_{\bm{p}}\big)$ to $\big(G'[\mathcal{V},\mathcal{H}',\mathcal{F}'],\omega_{\bm{p}'}\big)$ is a graph operation $h: \mathcal{H} \to \mathcal{H}'$ satisfying
\begin{enumerate}[(i)]
\item for any $(v,u) \in \mathcal{F}$, $(v,h(u)) \in \mathcal{F}'$;
\item for any $(v,u) \in \mathcal{F}$, 
\begin{equation}
\label{eq:losspre0}
    \tilde{l}(v,h(u)) \leq \tilde{l}(v,u);
\end{equation}
\item for any $(v,u) \in \mathcal{F}$ and $w \in \mathcal{W}$ (cf.~\eqref{p0}),
\begin{equation}
\label{eq:preserve0}
     p(w|h(u),v)=p(w|u,v);
\end{equation}
\item $\bm{p}'$ is naturally induced by $h$ from $\bm{p}$, to be precise, for any $(v,u') \in \mathcal{F}'$,
\begin{equation}
    \label{eq:defp'0}
    \omega_{\bm{p}'}(v,u') \triangleq p'(u'|v) = \sum_{\substack{u \in \mathcal{F}_v:\\ h(u) = u'}} p(u|v).
\end{equation}
\end{enumerate}
\end{definition}
\begin{remark}
\label{rem:welldefined}
For $\omega_{\bm{p}} \in \Omega(\mathcal{H},\mathcal{F})$, we have $\omega_{\bm{p}'} \in \Omega(\mathcal{H}',\mathcal{F}')$. The proof can be found in Appendix~\ref{sec:pfgbfo}. 
\end{remark}


\begin{lemma}
    \label{thm:bfo}
    For any subgraph  $G[\mathcal{V},\mathcal{H},\mathcal{F}]$ of $G[\mathcal{V},\mathcal{U},\mathcal{E}]$, if for each $\omega_{\bm{p}} \in \Omega(\mathcal{U},\mathcal{E})$, there exists an $\omega_{\bm{p}'} \in \Omega(\mathcal{H},\mathcal{F})$ so that we have a feasible contraction  from $\big(G[\mathcal{V},\mathcal{U},\mathcal{E}],\omega_{\bm{p}}\big)$ to $\big(G[\mathcal{V},\mathcal{H},\mathcal{F}],\omega_{\bm{p}'}\big)$, then 
\begin{equation}
\label{eq:pgraph'}
\begin{aligned}
    T(L) = &\min_{\omega_{\bm{p}'} \in \Omega(\mathcal{H},\mathcal{F})} \mathcal{O}(\bm{p}').
\end{aligned}
\end{equation}
\end{lemma}

The proof of \Cref{thm:bfo} can be found in Appendix~\ref{sec:pfbfo0}. 
Two special cases useful for the proof of~\Cref{thm:rdfci,thm:rdfmin,thm:ccfci,thm:ccfmin} are discussed in the following.
 
\subsection{The Case where $V$ and $W$ Admit a Joint Decomposition}
\label{subsec:com}
We consider the case where $V$ and $W$ {\it admit a joint decomposition}, which can further simplify the graph characterization. By generalizing the ideas of G\'{a}cs-K\"{o}rner-Witsenhausen common information,
we say $V$ and $W$ {\it admit a joint decomposition} if there exist partitions $\mathcal{V} = \cup_{k = 1}^K \mathcal{V}_k$ and $\mathcal{W} = \cup_{k = 1}^K \mathcal{W}_k$ for $K>1$ such that
for any $(v,w)$,
\begin{equation}
\label{eq:separation}
\exists u,\  p(w|u,v)>0 \Rightarrow \exists k, \ v \in \mathcal{V}_k, \ w \in \mathcal{W}_k.
\end{equation}

\begin{definition}
\label{def:genfeacon}
For a subgraph $G'[\mathcal{V},\mathcal{H}',\mathcal{F}']$ of $G[\mathcal{V},\mathcal{H},\mathcal{F}]$, 
a {\it generalized feasible contraction} from~$\big(G[\mathcal{V},\mathcal{H},\mathcal{F}],\omega_{\bm{p}}\big)$ to $\big(G'[\mathcal{V},\mathcal{H}',\mathcal{F}'],\omega_{\bm{p}'}\big)$ is a graph operation $\bm{h} = (h_k)_{k = 1}^K$, where $h_k: \mathcal{H} \to \mathcal{H}'$ satisfies
\begin{enumerate}[(i)]
\item $V$ and $W$ admit a joint decomposition (cf. \eqref{eq:separation});
\item for any $(v,u) \in \mathcal{F}$, 
$(v,h_k(u)) \in \mathcal{F}'$;
\item for any $v \in \mathcal{V}_k$ and $(v,u) \in \mathcal{F}$, 
\begin{equation}
\label{eq:losspre}
    \tilde{l}(v,h_k(u)) \leq \tilde{l}(v,u);
\end{equation}
\item  for any $v \in \mathcal{V}_k$, $(v,u) \in \mathcal{F}$ and $w \in \mathcal{W}_k$,
\begin{equation}
\label{eq:preserve}
    p(w|h_k(u),v)=p(w|u,v);
\end{equation}
\item for any $v \in \mathcal{V}_k$ and $(v,u') \in \mathcal{F}'$,
\begin{equation}
    \label{eq:defp'}
    \omega_{\bm{p}'}(v,u') \triangleq p'(u'|v) = \sum_{\substack{u \in \mathcal{F}_v:\\ h_k(u) = u'}} p(u|v).
\end{equation}
\end{enumerate}
\end{definition}

Then we have the following lemma proved in Appendix~\ref{sec:pfgbfo}. 
\begin{lemma}
    \label{thm:gbfo}
    For any subgraph  $G[\mathcal{V},\mathcal{H},\mathcal{F}]$ of $G[\mathcal{V},\mathcal{U},\mathcal{E}]$, if for each $\omega_{\bm{p}} \in \Omega(\mathcal{U},\mathcal{E})$, there exists an $\omega_{\bm{p}'} \in \Omega(\mathcal{H},\mathcal{F})$ so that we have a generalized feasible contraction  from $\big(G[\mathcal{V},\mathcal{U},\mathcal{E}],\omega_{\bm{p}}\big)$ to $\big(G[\mathcal{V},\mathcal{H},\mathcal{F}],\omega_{\bm{p}'}\big)$, then 
\begin{equation}
\label{eq:gpgraph'}
\begin{aligned}
    T(L) = &\min_{\omega_{\bm{p}'} \in \Omega(\mathcal{H},\mathcal{F})} \mathcal{O}(\bm{p}').
\end{aligned}
\end{equation}
\end{lemma}

Now we further investigate a special case of joint decomposition. Suppose $W = (\tilde{W},V')$,
\begin{equation}
\label{eq:psplit}
    p(\tilde{w},v'|u,v) = p(\tilde{w}|u,v,v')p(v'|v),
\end{equation}
and $V$ and $V'$ have a G\'{a}cs-K\"{o}rner-Witsenhausen common part (see~\Cref{def:GKW}).
Let $V_0$ be the GKW common part and $g_1: \mathcal{V} \to \{1,...,K\}$, $g_2: \mathcal{V}' \to  \{1,...,K\}$ satisfy $V_0 = g_1(V) = g_2(V')$.
Let $\mathcal{V}_k = g_1^{-1}(\{k\})$ and $\mathcal{W}_k = \tilde{\mathcal{W}} \times g_2^{-1}(\{k\})$, $k = 1,...,K$.
Then for any $k \neq k'$, $v \in \mathcal{V}_k$ and $w=(\tilde{w},v') \in \mathcal{W}_{k'}$, we have $g_1(v) = k$ and $g_2(v') = k'$, which implies $p(v,v')=0$. Then by~\eqref{eq:psplit}, $p(w|u,v) = 0$ for any $u$. 
So partitions $\mathcal{V} = \cup_{k = 1}^K \mathcal{V}_k$ and $\mathcal{W} = \cup_{k = 1}^K \mathcal{W}_k$ satisfy~\eqref{eq:separation}.
In other words, we have shown that in this case, $V$ and $W$ admit a joint decomposition under partitions $\mathcal{V} = \cup_{k = 1}^K \mathcal{V}_k$ and $\mathcal{W} = \cup_{k = 1}^K \mathcal{W}_k$ .

\subsection{The Minimum Loss Case}
\label{subsec:minloss}
Consider the minumum loss case that $L = L_{min}$.  For $\omega_{\bm{p}} \in \Omega(\mathcal{U},\mathcal{E})$, we can see that for any $(v,u) \in \mathcal{E}$, $\omega_{\bm{p}}(v,u)>0$ only if
\begin{equation}\label{eq:mincon}
    \tilde{l}(v,u) = \min_{u' \in \mathcal{E}_v}\left\{\tilde{l}(v,u')\right\}.
\end{equation}

To simplify the optimization problem in~\eqref{lem:pgraph}, we construct a subgraph $G_*[\mathcal{V},\mathcal{U}_*,\mathcal{E}_*]$ of $G[\mathcal{V},\mathcal{U},\mathcal{E}]$ by defining its edge set $\mathcal{E}_*$ and vertex set $\mathcal{U}_*$.
For any $e= (v,u) \in \mathcal{E}$, let $e \in \mathcal{E}_*$ if \eqref{eq:mincon} is satisfied.
By deleting vertices in $\mathcal{U}$ that are not adjacent to any edges in $\mathcal{E}_*$, we obtain $\mathcal{U}_* =\cup_{v \in \mathcal{V}}(\mathcal{E}_{*})_v$.
Then we obtain that $\omega_{\bm{p}} \in \Omega(\mathcal{U},\mathcal{E})$ is naturally equivalent to $\omega_{\bm{p}} \in \Omega(\mathcal{U}_*,\mathcal{E}_*)$, thus
we have the following result.

\begin{lemma}
\label{thm:minloss}
 $T(L_{min}) =\min_{\omega_{\bm{p}} \in \Omega(\mathcal{U}_*,\mathcal{E}_*)} \mathcal{O}(\bm{p})$.
\end{lemma}

\subsection{Proof of Theorem~\ref{thm:bfo}}
\label{sec:pfbfo0}

It is easy to see that the feasible contraction is a special case of the generalized feasible contraction, so~\Cref{thm:bfo} is an immediate corollary of \Cref{thm:gbfo} (note that the proof of \Cref{thm:gbfo} does not use \Cref{thm:bfo}).  

\subsection{Proof of Theorem \ref{thm:gbfo}}
\label{sec:pfgbfo}

We first prove the following lemma.

\begin{lemma}
\label{lem:bfo}
Suppose $\omega_{\bm{p}} \in \Omega(\mathcal{H},\mathcal{F})$ and there is a generalized feasible contraction from $\big(G[\mathcal{V},\mathcal{H},\mathcal{F}],\omega_{\bm{p}}\big)$ to $\big(G'[\mathcal{V},\mathcal{H}',\mathcal{F}'],\omega_{\bm{p}'}\big)$ , then $\omega_{\bm{p}'} \in \Omega(\mathcal{H}',\mathcal{F}')$ and $\mathcal{O}(\bm{p}') \leq \mathcal{O}(\bm{p})$.
\end{lemma}

\begin{IEEEproof}
We first show $\omega_{\bm{p}'} \in \Omega(\mathcal{H}',\mathcal{F}')$.
For any $v \in \mathcal{V}_k$, by~\eqref{eq:defp'} we have
\[
\begin{aligned}
    \sum_{u' \in \mathcal{F}_v'} p'(u'|v) &= \sum_{u' \in \mathcal{F}'_v} \sum_{\substack{u \in \mathcal{F}_v:\\ h_k(u) = u'}} p(u'|v) \\
    &=\sum_{u \in \mathcal{F}_v} p(u|v) = 1.
\end{aligned}
\]
By \eqref{eq:losspre}, we always have $\tilde{l}(v,u')\leq \tilde{l}(v,u) $ if $v \in \mathcal{V}_k$, $(v,u') \in \mathcal{F}'$ and $h_k(u) = u'$. Then by \eqref{eq:separation} and \eqref{eq:defp'},
\begin{align*}
    &Loss(\bm{p'}) = \sum_{\substack{k,v \in \mathcal{V}_k,\\u'\in \mathcal{F}'_v}}p(v)p'(u'|v)\tilde{l}(v,u')\\
    \leq & \sum_{\substack{k,v \in \mathcal{V}_k,\\u'\in \mathcal{F}'_v}}\sum_{\substack{u \in \mathcal{F}_v:\\ h_k(u) = u'}}p(v)p(u|v)\tilde{l}(v,u)= Loss(\bm{p}).
\end{align*}
So $\omega_{\bm{p}'} \in \Omega(\mathcal{H}',\mathcal{F}')$.

It remains to show $\mathcal{O}(\bm{p}') \leq \mathcal{O}(\bm{p})$.
For any $w \in \mathcal{W}_k$, by \eqref{eq:separation}, \eqref{eq:preserve} and \eqref{eq:defp'} we have
\begin{align*}
    &p'(u',w)  = \sum_{\substack{v \in \mathcal{V}_k \cap (\mathcal{F}')^{u'}}} p(v)p'(u'|v)p(w|u',v) \\
    = &\sum_{\substack{v \in \mathcal{V}_k \cap (\mathcal{F}')^{u'} }} \sum_{\substack{u \in \mathcal{F}_v:\\ h_k(u) = u'}} p(v)p(u|v)p(w|u,v) \\
    = &\sum_{\substack{u:h_k(u) = u'}} \sum_{v \in \mathcal{V}_k \cap \mathcal{F}^u} p(v)p(u|v)p(w|u,v) \\
    = &\sum_{u:h_k(u) = u'} p(u,w),
\end{align*}
which implies $p'(u'|w) = \sum_{u:h_k(u) = u'} p(u|w)$.
Then by \eqref{eq:separation}, \eqref{eq:preserve} and \eqref{eq:defp'} we have
\begin{align*}
&\mathcal{O}(\bm{p'})-\mathcal{O}(\bm{p})\\
=&\sum_{\substack{k,v\in \mathcal{V}_k,\\ w \in \mathcal{W}_k,\\u' \in \mathcal{F}'_v}}\sum_{\substack{u \in \mathcal{F}_v:\\h_k(u)=u'}}p(v)p(u|v)p(w|u',v)\log \frac{p(u|w)p'(u'|v)}{p(u|v)p'(u'|w)}\\
\leq& \sum_{\substack{k,v\in \mathcal{V}_k,\\ w \in \mathcal{W}_k,\\u' \in \mathcal{F}'_v}}\sum_{\substack{u \in \mathcal{F}_v:\\h_k(u)=u'}}p(v)p(u|v)p(w|u',v)\frac{p(u|w)p'(u'|v)}{p(u|v)p'(u'|w)}-1\\
=&\sum_{\substack{k,v\in \mathcal{V}_k,\\ w \in \mathcal{W}_k,\\u' \in \mathcal{F}'_v}}p(v)p'(u'|w)p(w|u',v)-1=0,
\end{align*}
which completes the proof. 
\end{IEEEproof}

By \eqref{lem:pgraph}, it suffices to show $\min_{\omega_{\bm{p}} \in \Omega(\mathcal{U},\mathcal{E})} \mathcal{O}(\bm{p}) = \min_{\omega_{\bm{p}'} \in \Omega(\mathcal{H},\mathcal{F})} \mathcal{O}(\bm{p}')$. Note that $\mathcal{H} \subseteq \mathcal{U}$ and $\mathcal{F} \subseteq \mathcal{E}$, then any $\omega_{\bm{p}} \in \Omega(\mathcal{H},\mathcal{F})$ is naturally an element in $\Omega(\mathcal{U},\mathcal{E})$. This gives $\min_{\omega_{\bm{p}} \in \Omega(\mathcal{U},\mathcal{E})} \mathcal{O}(\bm{p}) \leq \min_{\omega_{\bm{p}'} \in \Omega(\mathcal{H},\mathcal{F})} \mathcal{O}(\bm{p}')$. The other direction is immediate from \Cref{lem:bfo}.

\section{Proof of Lemma~\ref{lem:rdf}}
\label{sec:pfrdf}
Recall that we have
\begin{equation*}
R(D) = \min_{\substack{p(u|s_1):\  \exists g, \\  \mathbb{E}[d(f(S_1,S_2),g(U,S_2))] \leq D}}I(U;S_1)-I(U;S_2).
\end{equation*}

By assignments in~\eqref{eq:assrdf}, we need to show 
\begin{align*}
    \min_{\substack{p(u|s_1):\  \exists g, \\  \mathbb{E}[d(f(S_1,S_2),g(U,S_2))] \leq D}}I(U;S_1)-I(U;S_2) 
    \\
    = \min_{\omega_{\bm{p}} \in \Omega(\mathcal{U},\mathcal{E})} \mathcal{O}(\bm{p}).
\end{align*}
First we interpret the left hand side from a graph point of view. 
For each feasible solution $(\bm{p},g)$, denote the alphabet of $U$ by $\mathcal{U}_g$ and define $l_g(s_1,u,s_2) = d(f(s_1,s_2),g(u,s_2))$. We can construct a bipartite graph $G_g[\mathcal{S}_1,\mathcal{U}_g,\mathcal{E}_g]$ with $\mathcal{E}_g = \mathcal{S}_1 \times \mathcal{U}_g$.
By $\mathbb{E}[d(f(S_1,S_2),g(U,S_2))] \leq D$, $L = D$ and the definition of $l_g$, 
$\omega_{\bm{p}} \in \Omega(\mathcal{U}_g,\mathcal{E}_g)$. So the left hand side can be alternatively characterized by $\min_{g,\omega_{\bm{p}} \in \Omega(\mathcal{U}_g,\mathcal{E}_g)} \mathcal{O}(\bm{p})$, and then it suffices to prove 
\begin{equation}
\label{eq:gpgraph}
    \min_{g,\omega_{\bm{p}} \in \Omega(\mathcal{U}_g,\mathcal{E}_g)} \mathcal{O}(\bm{p}) = \min_{\omega_{\bm{p}} \in \Omega(\mathcal{U},\mathcal{E})} \mathcal{O}(\bm{p}).
\end{equation}

We can always find an optimal $(g, \bm{p})$ satisfying that for any $u,u' \in \mathcal{U}_g$, there exists an $s_2$ such that 
\begin{equation}
\label{eq:assump}
     g(u,s_2) \neq g(u',s_2).
\end{equation} 
For any optimal $(g, \bm{p})$, if the above constraint is not satisfied, we apply a feasible contraction on $\big(G_g[\mathcal{S}_1,\mathcal{U}_g,\mathcal{E}_g],\omega_{\bm{p}}\big)$ as follows. We construct a partition $\mathcal{U}_g= \cup_i \mathcal{U}_g^i$ with respect to the value of $g(u,s_2)$, that is, $u$ and $u'$ are in the same $\mathcal{U}_g^i$ if $g(u,s_2) = g(u',s_2)$ $\forall s_2 \in \mathcal{S}_2$.  Then choose one representative $u_i$ from each $U_g^i$ and define the feasible contraction by letting $h(u) = u_i $ if $u \in \mathcal{U}_g^i$. The constraints in~\Cref{def:feacon} can be easily verified. Denote the weighted graph after the contraction by $(G_g'[\mathcal{S}_1,\mathcal{U}_g',\mathcal{E}_g'],\omega_{\bm{p}'})$, where $\mathcal{U}_g' = \cup_i \{u_i\}$, $\mathcal{E}_g' = \mathcal{S}_1\times \mathcal{U}_g'$ and $\omega_{\bm{p}'}$ is obtained accordingly.
By~\Cref{lem:bfo}, $\omega_{\bm{p}'} \in \Omega(\mathcal{U}_g',\mathcal{E}_g')$ and $\mathcal{O}(\bm{p}') \leq \mathcal{O}(\bm{p})$, which implies the optimality of $(g,\bm{p}')$.
Finally, by the structure of~$\mathcal{U}_g'$,
we see that for any $u,u' \in \mathcal{U}_g'$, there exists an $s_2$ satisfying~\eqref{eq:assump}.

With the above assumption~\eqref{eq:assump}, any $G_g[\mathcal{S}_1,\mathcal{U}_g,\mathcal{E}_g]$ can be viewed as a subgraph of $G[\mathcal{V},\mathcal{U},\mathcal{E}]$ by applying an injective mapping $\mathcal{U}_g \to \mathcal{U}$ that maps $u \in \mathcal{U}_g$ to $(g(u,s_2))_{s_2 \in \mathcal{S}_2} \in \mathcal{U}$. 
Then we have
\[
\min_{\omega_{\bm{p}} \in \Omega(\mathcal{U},\mathcal{E})} \mathcal{O}(\bm{p}) \leq
\min_{g,\omega_{\bm{p}} \in \Omega(\mathcal{U}_g,\mathcal{E}_g)} \mathcal{O}(\bm{p}).
\]

Finally, let $g(u,s_2) = \hat{z}_{s_2}$ for each $u = (\hat{z}_{s_2})_{s_2 \in \mathcal{S}_2} \in \mathcal{U}$ and $s_2 \in \mathcal{S}_2$. 
By~\eqref{eq:assrdf} we have $\mathcal{U}_g = \mathcal{U}$ 
 and $\mathcal{E}_g = \mathcal{E}$. 
Then each $\omega_{\bm{p}} \in \Omega(\mathcal{U},\mathcal{E})$ is also in $\Omega(\mathcal{U}_g,\mathcal{E}_g)$ and hence
\[
\min_{g,\omega_{\bm{p}} \in \Omega(\mathcal{U}_g,\mathcal{E}_g)} \mathcal{O}(\bm{p})\leq
\min_{\omega_{\bm{p}} \in \Omega(\mathcal{U},\mathcal{E})} \mathcal{O}(\bm{p}).
\]
This completes the proof.

\section{Proof of Lemma~\ref{lem:ccf}}
\label{sec:pfccf}
Similarly by the results in~\cite{Cover2002}, the capacity-cost function for the channel can be written as
\begin{equation}\label{pccf}
C(B) = \max_{\substack{p(u|s_1): \ \exists g, \\ \mathbb{E}[b(g(U,S_1),S_1,S_2)] \leq B}}I(U;Y,S_2)-I(U;S_1). 
\end{equation}
Then the rest of the proof is similar to~\Cref{lem:rdf} and omitted.

\section{Proof of Lemma~\ref{lem:crdf}}
\label{sec:pfcrdf}
Recall that by~\eqref{pcrdf} we have
\begin{equation*}
R(D) = \min_{\substack{p(u|s_1):\  \exists g, \\  \mathbb{E}[d(f(S_1,S_2),g(U,S_2))] \leq D}}I(U;S_1).
\end{equation*}

By assignments in~\eqref{eq:asscrdf}, we need to show 
\begin{align*}
    \min_{\substack{p(u|s_1):\  \exists g, \\  \mathbb{E}[d(f(S_1,S_2),g(U,S_2))] \leq D}}I(U;S_1) 
    = \min_{\omega_{\bm{p}} \in \Omega(\mathcal{U},\mathcal{E})} \mathcal{O}(\bm{p}).
\end{align*}
Similar to the proof of~\Cref{lem:rdf} in Appendix~\ref{sec:pfrdf}, we can interpret the left hand side from a graph point of view. 
For each feasible solution $(\bm{p},g)$, denote the alphabet of $U$ by $\mathcal{U}_g$ and define $l_g(s_1,u) = \sum_{s_2} p(s_2|s_1)d(f(s_1,s_2),g(u,s_2))$. We can construct a bipartite graph $G_g[\mathcal{S}_1,\mathcal{U}_g,\mathcal{E}_g]$ with $\mathcal{E}_g = \mathcal{S}_1 \times \mathcal{U}_g$.
By $\mathbb{E}[d(f(S_1,S_2),g(U,S_2))] \leq D$, $L = D$ and the definition of $l_g$, 
$\omega_{\bm{p}} \in \Omega(\mathcal{U}_g,\mathcal{E}_g)$. So the left hand side can be alternatively characterized by $\min_{g,\omega_{\bm{p}} \in \Omega(\mathcal{U}_g,\mathcal{E}_g)} \mathcal{O}(\bm{p})$, and then it suffices to prove 
\begin{equation*}
    \min_{g,\omega_{\bm{p}} \in \Omega(\mathcal{U}_g,\mathcal{E}_g)} \mathcal{O}(\bm{p}) = \min_{\omega_{\bm{p}} \in \Omega(\mathcal{U},\mathcal{E})} \mathcal{O}(\bm{p}).
\end{equation*}

The remaining proof is the same as that for~\Cref{lem:rdf} in Appendix~\ref{sec:pfrdf}.

\section{Proof of Lemma \ref{lem:con opt}}
\label{sec:pfcon opt}

We first prove the problems~\eqref{p0} and~\eqref{p2} have the same optimal value. Denote the optimal value of~\eqref{p2} by $T'(L)$. Then it is immediate that $T'(L) \leq T(L)$. We now prove the other direction. 
Let \[
GD_{\bm{q}_{U|V}}(\bm{r}_1||\bm{r}_2) = \sum_{u,w} q(w)r_1(u|w) \log \frac{r_1(u|w)}{r_2(u|w)},
\]
where $q(w) \triangleq \sum_{(v,u)\in \mathcal{E}}p(v)q(u|v)p(w|u,v)$.
Note that $GD_{\bm{q}}(\bm{r}_1||\bm{r}_2) \geq 0, \forall \bm{r}_1, \bm{r}_2$.
We can verify for any $(\bm{q},\bm{r})$ that
\begin{equation}
\begin{aligned}
    &GD_{\mathcal{E}}(\bm{q}_{U|V}||\bm{r}_{U|W}) \\
    = & GD_{\mathcal{E}}(\bm{q}_{U|V}||\bm{q}_{U|W})+ GD_{\bm{q}_{U|V}}(\bm{q}_{U|W}||\bm{r}_{U|W}) 
    \\
    \geq &GD_{\mathcal{E}}(\bm{q}_{U|V}||\bm{q}_{U|W}) \geq T(L).
\end{aligned}
\end{equation}
So $T(L) \leq T'(L)$, which proves $T(L) = T'(L)$.

Next, we show \eqref{p2} is a convex optimization problem.  Since the constraints are linear, we only need to verify the convexity of the objective function $GD_{\mathcal{E}}(\bm{q}||\bm{r})$. 
Let $(\bm{q}_1,\bm{r}_1)$ and $(\bm{q}_2,\bm{r}_2)$ be two feasible solutions and $0<\alpha < 1$. By the log-sum inequality, we have 
\[
\begin{aligned}
&[(1-\alpha)q_1(u|v)+\alpha q_2(u|v)]\log \frac{(1-\alpha)q_1(u|v)+\alpha q_2(u|v)}{(1-\alpha)r_1(u|w)+\alpha r_2(u|w)}
\\
&\leq  (1-\alpha)q_1(u|v)\log \frac{q_1(u|v)}{r_1(u|w)} + \alpha q_2(u|v)\log \frac{q_2(u|v)}{r_2(u|w)}.
\end{aligned}
\]
Multiplying both sides by $p(v)p(w|u,v)$ and taking the sum over all $(u,v) \in \mathcal{E}$ and $w \in \mathcal{W}$, we have
\begin{align*}
GD_{\mathcal{E}}((1-\alpha)\bm{q}_1+\alpha \bm{q}_2||(1-\alpha)\bm{r}_1+\alpha \bm{r}_2) \\
\leq (1-\alpha)GD_{\mathcal{E}}(\bm{q}_1||\bm{r}_1)+\alpha 
 GD_{\mathcal{E}}(\bm{q}_2||\bm{r}_2).
\end{align*}
 
Finally, the convexity of the objective function $\mathcal{O}(\bm{p})$ of~\eqref{p0} is immediate from the convexity of $GD_{\mathcal{E}}(\bm{q}||\bm{r})$ and
 the relation $\mathcal{O}(\bm{q}) =  \min_{\bm{r}} GD_{\mathcal{E}}(\bm{q}||\bm{r})$.

\section{Proof of Lemma \ref{lemma:T(L)}}
\label{sec:pfTL}

Since the feasible region of~\eqref{p0} expands as $L$ increases, the optimal value $T(L)$ is non-increasing in $D$.

Let $\bm{q}_1$ and $\bm{q}_2$ be two feasible solutions of~\eqref{p0} for $L_1$ and $L_2$, respectively.
Let $\alpha \in (0,1)$.
Since the constraints are linear, $(1-\alpha)\bm{q}_1+ \alpha \bm{q}_2$ is a feasible solution for $(1-\alpha)L_1+\alpha L_2$.
Then by the convexity of $\mathcal{O}(\bm{q})$ in \Cref{lem:con opt} and the definition of $T(L)$, we have for any $\bm{q}_1$ and $\bm{q}_2$ that
\begin{align*}
T((1-\alpha)L_1+\alpha L_2) \leq \mathcal{O}((1-\alpha)\bm{q}_1+ \alpha \bm{q}_2) \\
\leq (1-\alpha)\mathcal{O}(\bm{q}_1)+ \alpha \mathcal{O}(\bm{q}_2),
\end{align*}
which implies
\[
T((1-\alpha)L_1+\alpha L_2) \leq (1-\alpha)T(L_1)+\alpha T(L_2).
\]
It gives the convexity of $T(L)$.

\section{Proof of Theorem \ref{thm:rdfci}}
\label{sec:pfrdfci}
Since~\eqref{eq:assrdfd} satisfies~\eqref{eq:psplit}, following the discussion at the end of~\Cref{subsec:com}, we see that $S_1$ and $S_2$ admit a joint decomposition.
Then by~\Cref{def:genfeacon}, we can construct a contraction on the graph so that~\Cref{thm:gbfo} can be applied to simplify the graph characterization.

We first add a point $\infty$ with $d(z, \infty) = +\infty, \forall z \in \mathcal{Z}$ to $\hat{\mathcal{Z}}$ and the rate-distortion function does not change. 
We apply a generalized feasible contraction on the characteristic bipartite graph $G[\mathcal{V},\mathcal{U},\mathcal{E}]$ as follows. 
For any $k = 1,...,K$, define $H_{k}(s_2,\hat{z})=\hat{z}$ if $g_2(s_2) = k$ and $H_{k}(s_2,\hat{z})=\infty$ otherwise. Define $h_{k}((\hat{z}_{s_2})_{s_2 \in \mathcal{S}_2}) = (H_{k}(s_2,\hat{z}_{s_2}))_{s_2 \in \mathcal{S}_2}$.
Denote the graph after the contraction by $G'[\mathcal{V},\mathcal{U}',\mathcal{E}']$, where 
$\mathcal{U}' = \cup_{k =1}^K \mathcal{U}_k'$, 
$$\mathcal{U}_k' = \{(\hat{z}_{s_2})_{s_2 \in \mathcal{S}_2}| \hat{z}_{s_2} = \infty, \forall s_2 \notin \mathcal{S}_{2k}\}$$
and
$$\mathcal{E}' = \cup_{k = 1}^K \mathcal{S}_{1k} \times \mathcal{U}_k'.$$

Then by~\eqref{eq:assrdfd}, for any $s_1 \in \mathcal{S}_{1k}$ and $s_2 \in \mathcal{S}_{2k}$,
\begin{equation*}
     p(s_2|h_k(u),s_1) =p(s_2|s_1) =p(s_2|u,s_1),
\end{equation*}
and for any $s_1 \in \mathcal{S}_{1k}$,
\begin{equation*}
\begin{aligned}
    \tilde{l}(s_1,h_k(u)) = \sum_{s_2 \in \mathcal{S}_{2k}}d(f(s_1,s_2),H_k(s_2,\hat{z}_{s_2}))p(s_2|s_1) \\
    =\sum_{s_2 \in \mathcal{S}_{2k}}d(f(s_1,s_2),\hat{z}_{s_2})p(s_2|s_1)
    = \tilde{l}(s_1,u),
\end{aligned}
\end{equation*}
so that all the constraints in~\Cref{def:genfeacon} are satisfied. Hence $\bm{h} = (h_k)_{k =1}^K$ is a generalized feasible contraction from $G[\mathcal{V},\mathcal{U},\mathcal{E}]$ to $G'[\mathcal{V},\mathcal{U}',\mathcal{E}']$.

By \Cref{thm:gbfo} we have $R(D) = \min_{\omega_{\bm{p}} \in \Omega(\mathcal{U}',\mathcal{E}')} \mathcal{O}(\bm{p})$, which is equal to~\eqref{eq:rdfci} except that $\infty$ is in $\hat{\mathcal{Z}}$.
Next, we remove the vertex $(\infty)_{s_2 \in \mathcal{S}_2} \in \mathcal{U}'$ and edges adjacent to it from the graph $G'[\mathcal{V},\mathcal{U}',\mathcal{E}']$ without changing the optimal value of the graph characterization. 
Then $\mathcal{U}'-\{(\infty)_{s_2 \in \mathcal{S}_2}\}= \cup_{k =1}^K \mathcal{U}_k'-\{(\infty)_{s_2 \in \mathcal{S}_2}\}$ is a partition.
Vertices in $\mathcal{S}_{1k}$ are exactly connected to vertices in~$\mathcal{U}_k'-\{(\infty)_{s_2 \in \mathcal{S}_2}\}$ that can be viewed as $(\hat{z}_{s_2})_{s_2 \in \mathcal{S}_{2k}}$ by eliminating redundant components $\infty$.
Moreover, such vertices $(\hat{z}_{s_2})_{s_2 \in \mathcal{S}_{2k}}$ with $\hat{z}_{s_2} = \infty$ for some $s_2 \in \mathcal{S}_{2k}$ can also be deleted and we complete the proof.

\section{Proof of Theorem~\ref{thm:rdfmin}}
\label{sec:pfrdfmin}
In light of~\Cref{thm:minloss}, we have a simplified rate-distortion function $R(D_{min}) = \min_{\omega_{\bm{p}} \in \Omega(\mathcal{U}_*,\mathcal{E}_*)} \mathcal{O}(\bm{p})$, where $\mathcal{U}_*$ and $\mathcal{E}_*$ are defined in \Cref{subsec:minloss}. 

By the definition of $\Gamma_m(\mathcal{S}_1)$, for each $u \in \mathcal{U}_*$, we can always find a $\mathcal{C}_u \in \Gamma_m(\mathcal{S}_1)$ such that $\mathcal{E}_*^{u} \subseteq \mathcal{C}_u$. Define a map $\mathscr{C}: \mathcal{U}_* \to \Gamma_m(\mathcal{S}_1)$ with $\mathscr{C}(u) = \mathcal{C}_u$ and then $\mathcal{U}_* = \cup_{\mathcal{C} \in \Gamma_m(\mathcal{S}_1)}\mathscr{C}^{-1}(\mathcal{C})$ is a partition.
Note that for any $\mathcal{C} \in \Gamma_m(\mathcal{S}_1)$, by the definition of $\Gamma_m(\mathcal{S}_1)$ there exists some $u_{\mathcal{C}} \in \mathcal{U}$ such that $\mathcal{C} = \mathcal{E}_*^{u_{\mathcal{C}}}$, then we have $u_{\mathcal{C}} \in \mathscr{C}^{-1}(\mathcal{C})$.
We can perform a feasible contraction on the bipartite graph $G_{*}[\mathcal{S}_1,\mathcal{U}_*,\mathcal{E}_*]$ by defining $h(u) = u_{\mathcal{C}}$ if $u \in \mathscr{C}^{-1}(\mathcal{C})$.
Denote the graph after the contraction by $G_{**}[\mathcal{V},\mathcal{U}_{**},\mathcal{E}_{**}]$, where $\mathcal{U}_{**} = \{u_{\mathcal{C}}, \mathcal{C} \in \Gamma_m(\mathcal{S}_1) \} \subseteq \mathcal{U}_*$ and
$$\mathcal{E}_{**} = \cup_{\mathcal{C} \in \Gamma_m(\mathcal{S}_1)} \mathcal{C} \times \{u_{\mathcal{C}}\} \subseteq \mathcal{E}_*.$$

For any $u \in \mathscr{C}^{-1}(\mathcal{C})$, we have $\mathcal{E}_*^{u} \subseteq \mathcal{C} = \mathcal{E}_*^{u_{\mathcal{C}}}$,
and hence $\tilde{l}(s_1,h(u)) = \min_{u' \in \mathcal{U}}\tilde{l}(s_1,u') = \tilde{l}(s_1,u)$ for any $(s_1,u) \in \mathcal{E}_*$.
Also by~\eqref{eq:assrdfd},
so all the constraints in~\Cref{def:feacon} are satisfied. Hence $h$ is a feasible contraction from $G_{*}[\mathcal{V},\mathcal{U}_*,\mathcal{E}_*]$ to $G_{**}[\mathcal{V},\mathcal{U}_{**},\mathcal{E}_{**}]$.
By exploiting~\Cref{thm:bfo}, we complete the proof.

\section{Proof of Theorem~\ref{thm:ccfmin}}
\label{sec:pfccfmin}
We start from the result by \Cref{thm:minloss}, $C(B) = -\min_{\omega_{\bm{p}} \in \Omega(\mathcal{U}_*,\mathcal{E}_*)} \mathcal{O}(\bm{p})$, and apply a feasible contraction as follows.
Fix an arbitrary $x_{s_1}^* \in \mathcal{X}_{s_1}$, then define $H(s_1, x) = x$ if $x \in \mathcal{X}_{s_1}$ and $H(s_1,x) = x_{s_1}^*$ otherwise. For any $u = (x_{s_1})_{s_1 \in \mathcal{S}_1}$, define $h(u) = (H(s_1,x_{s_1}))_{s_1 \in \mathcal{S}_1}$.
Denote the graph after the contraction by $G_{**}[\mathcal{V},\mathcal{U}_{**},\mathcal{E}_{**}]$, where $\mathcal{U}_{**}  = \prod_{s_1' \in \mathcal{S}_1}\mathcal{X}_{s_1'} \subseteq \mathcal{U}_*$ and $\mathcal{E}_{**} = \mathcal{S}_1 \times \mathcal{U}_{**}$.

For any $(s_1,u) \in \mathcal{E}_{*}$, $u = (x_{s_1'})_{s_1' \in \mathcal{S}_1}$, by~\eqref{eq:Bmincon} we have $x_{s_1} \in \mathcal{X}_{s_1}$.
Then by~\eqref{eq:lformula} and~\eqref{eq:defXs1} we have $$\tilde{l}(s_1,h(u)) = \min_{u \in \mathcal{U}}\left\{\tilde{l}(s_1,u)\right\} = \tilde{l}(s_1,u).$$
Also by~\eqref{eq:assccfd},
\[
\begin{aligned}
p(y,s_2|h(u),s_1) = p(s_2|s_1)p(y|H(s_1,x_{s_1}),s_1,s_2)
\\= p(s_2|s_1)p(y|x_{s_1},s_1,s_2)= p(y,s_2|u,s_1).
\end{aligned}
\]
So the constraints in~\Cref{def:feacon} is satisfied and $h$ is a feasible contraction from $G_*[\mathcal{V},\mathcal{U}_*,\mathcal{E}_*]$ to $G_{**}[\mathcal{V},\mathcal{U}_{**},\mathcal{E}_{**}]$.
By \Cref{thm:bfo}, we complete the proof.

\section{Proof of the Partial Minimization Process in Definition~\ref{def:parmin}}
\label{subsec:partialmin}

The partial minimization process is immediate given the following lemma, by noting that $GD_{\mathcal{E}}(\bm{q}_1||\bm{q}_2)$ and $GD_{\bm{q}}(\bm{r}_1||\bm{r}_2)$ defined below are linear combinations of K-L divergence and are  always non-negative. 

\begin{lemma}
\label{lem:Fs1}
We have the following identities for $\bm{q}$ with $\mathrm{supp}(\bm{q}) \subseteq \mathcal{E}$, where the generalized K-L divergence for $q(u|v)$ and $r(u|w)$ are naturally defined as
\[
    GD_{\mathcal{E}}(\bm{q}_1||\bm{q}_2) = \sum_{(v,u)\in \mathcal{E}} p(v)q_1(u|v) \log \frac{q_1(u|v)}{q_2(u|v)}
\]
and 
\[
GD_{\bm{q}}(\bm{r}_1||\bm{r}_2) = \sum_{u,w} q(w)r_1(u|w) \log \frac{r_1(u|w)}{r_2(u|w)},
\]
where $q(w) \triangleq \sum_{(v,u) \in \mathcal{E}}p(v)q(u|v)p(w|u,v)$.
\begin{align}
&F_s(\bm{q},\bm{r}) = F_s(\bm{q}_s^*(\bm{r}),\bm{r})+ GD_{\mathcal{E}}(\bm{q}||\bm{q}_s^*(\bm{r})),\label{eq:Fsq}
\\
&F_s(\bm{q},\bm{r}) = F_s(\bm{q},\bm{r}^*(\bm{q}))+ GD_{\bm{q}}(\bm{r}^*(\bm{q})||\bm{r}).\label{eq:Fsr}
\end{align}
\end{lemma}

\begin{IEEEproof}
First we show~\eqref{eq:Fsq}.
By~\eqref{eq:alterq}, we can compute that 
\begin{equation*}
\begin{aligned}
    & F_s(\bm{q}_s^*(\bm{r}),\bm{r})+ GD_{\mathcal{E}}(\bm{q}||\bm{q}_s^*(\bm{r}))
    \\
    =&-s\sum_{(v,u)\in \mathcal{E}} \! p(v)q_s^*(\bm{r})(u|v) \tilde{l}(v,u)
    \\
    +&\sum_{(v,u)\in \mathcal{E},w'} \! p(v)q_s^*(\bm{r})(u|v)p(w'|u,v) \log r(u|w')
    \\
    -&\sum_{v} \! p(v) \log  \sum_{u' \in \mathcal{E}_v} e^{-s \tilde{l}(v,u')}\prod_{w'} r(u'|w')^{p(w'|u',v)}
    \\
    -&\sum_{(v,u)\in \mathcal{E},w} \! p(v)q_s^*(\bm{r})(u|v)p(w|u,v) \log r(u|w)
    \\
    +& s \sum_{(v,u)\in \mathcal{E}} p(v)q_s^*(\bm{r})(u|v) \tilde{l}(v,u)
    \\
    +&\sum_{(v,u)\in \mathcal{E},w} \! p(v)q(u|v)p(w|u,v) \log q(u|v)
    \\
    +&s\sum_{(v,u)\in \mathcal{E}} \! p(v)q(u|v) \tilde{l}(v,u)
    \\
    -&\sum_{(v,u)\in \mathcal{E},w'} \! p(v)q(u|v)p(w'|u,v) \log r(u|w')
    \\
    +&\sum_{v} \! p(v) \log  \sum_{u' \in \mathcal{E}_v} e^{-s \tilde{l}(v,u')}\prod_{w'} r(u'|w')^{p(w'|u',v)}
    \\
    =&\sum_{(v,u)\in \mathcal{E},w} \! p(v)q(u|v)p(w|u,v) \log \frac{q(u|v)}{r(u|w)}
    \\
    +& s \sum_{(v,u)\in \mathcal{E}} p(v)q(u|v) \tilde{l}(v,u) 
    = F_s(\bm{q},\bm{r}).
\end{aligned} 
\end{equation*}

Then we proceed to show~\eqref{eq:Fsr}. By~\eqref{eq:alterr} and the definition of $q(w)$, we have
\begin{equation*}
\begin{aligned}
    &F_s(\bm{q},\bm{r}^*(\bm{q}))+ GD_{\bm{q}}(\bm{r}^*(\bm{q})||\bm{r})
    \\
    =&\sum_{(v,u)\in \mathcal{E},w} \! p(v)q(u|v)p(w|u,v) \log q(u|v)
    \\
    -&\sum_{(v,u)\in \mathcal{E},w} \! p(v)q(u|v)p(w|u,v) \log r^*(\bm{q})(u|w)
    \\
    +& s \sum_{(v,u)\in \mathcal{E}} p(v)q(u|v) \tilde{l}(v,u)
    \\
    +&\sum_{u,w} q(w)r^*(\bm{q})(u|w) \log \frac{r^*(\bm{q})(u|w)}{r(u|w)}
    \\
    =&\sum_{(v,u)\in \mathcal{E},w} \! p(v)q(u|v)p(w|u,v) \log \frac{q(u|v)}{r(u|w)}
    \\
    +& s \sum_{(v,u)\in \mathcal{E}} p(v)q(u|v) \tilde{l}(v,u) 
    = F_s(\bm{q},\bm{r}),
\end{aligned}
\end{equation*}
completing the proof. 
\end{IEEEproof}

\section{Proof of Theorem \ref{thm:convmain}}
\label{subsec:pfconvmain}

Let $(\bm{q}^0,\bm{r^0}) \in \arg \min_{\bm{q},\bm{r}}F_{s^*}(\bm{q},\bm{r})$.
For an algorithm with the iteration step~\eqref{eq:generalalter}, define a discriminant to be
\begin{equation}
    \Delta_n(\bm{q}^0) \triangleq (s^{(n)}-s^*)(Loss(\bm{q}^{(n)})-Loss(\bm{q}^0)). 
\end{equation}

To handle two cases (i.e. $L \in (L_{min}, L_{Max})$ or $L= L_{min}, L_{Max}$) in~\Cref{alg3} together, first we show that the condition $\Delta_n(\bm{q}^0) \geq~0$  can guarantee the convergency. We summarize the results in the following lemma, which is proved in~\Cref{subsec:pfthm:conv}. 

\begin{lemma}\label{thm:conv}
\begin{enumerate}[1)]
\item If there exists some $(\bm{q}^0,\bm{r^0}) \in \arg \min_{\bm{q},\bm{r}}F_{s^*}(\bm{q},\bm{r})$ such that $ \Delta_n(\bm{q}^0) \geq 0, \forall n$, then we have
$$\lim_{n \to \infty}F_{s^*}(\bm{q}^{(n+1)},\bm{r}^{(n)})=\min_{\bm{q},\bm{r}}F_{s^*}(\bm{q},\bm{r}).$$
Also, the optimal value achieved by the first $k$ iterations is characterized by
\begin{equation}
\label{eq:generalrate}
\begin{aligned}
    \min_{1 \leq k \leq n}F_{s^*}(\bm{q}^{(k+1)},\bm{r}^{(k)})-\min_{\bm{q},\bm{r}}F_{s^*}(\bm{q},\bm{r}) = \\ O\left(\frac{\log |\mathcal{U}|}{n}\right).
\end{aligned}
\end{equation}

\item Suppose $\bm{q}$ satisfies $\mathrm{supp}(\bm{q}) \subseteq \mathcal{E}$ and fix (P) to be one of the following two problems:
\begin{subequations}
\label{pall}
\begin{align}
&\min_{\bm{q},\bm{r}} F_{s^*}(\bm{q},\bm{r}),
\label{palla}
\\
&\min_{\bm{q},\bm{r}:Loss(\bm{q}) \leq L}GD_{\mathcal{E}}(\bm{q}||\bm{r}).
\label{pallc}
\end{align}
\end{subequations}

For any optimal solution $(\bm{q}^0,\bm{r^0})$ of (P),
if
$ \Delta_n(\bm{q}^0) \geq 0$ and $\bm{q}^{(n)}$ satisfies the loss constraint in (P), $\forall n$,
then we further have $(\bm{q}^{(n+1)},\bm{r}^{(n)})$ converge to an optimal solution of (P).
Moreover, the convergent rate for the objective function of~\eqref{palla} is $O\left(\frac{\log|\mathcal{U}|}{n}\right)$.
\end{enumerate}
\end{lemma}

To show~\Cref{thm:convmain}, consider the two cases in~\Cref{alg3} respectively. 
The case for $L = L_{min}$ or $L = L_{Max}$ is transformed into an equivalent form~\eqref{palla} with $s^* = 0$.
In this case, \Cref{alg3} is equivalent to a BA type algorithm for $s^{(n)} = s' = 0, \forall n$.
Hence we always have $\Delta_n(\bm{q}^0) = 0$, and \Cref{thm:conv} gives the desired $O(\frac{\log|\mathcal{U}|}{n})$ convergence rate.

For $L \in (L_{min},L_{Max})$, we have the following lemma shown in~\Cref{subsec:pflem:greaterKKT}.
\begin{lemma}
\label{lem:greaterKKT}
The choice of $s^{(n)}$ by \Cref{strategy1} satisfies the convergence condition $\Delta_n(\bm{q}^0) \geq 0$ for any optimal solution $(\bm{q}^{0},\bm{r}^{0})$ of the problem~\eqref{pallc}.
\end{lemma}
Note that~\eqref{pallc} is a simple repetition of our main goal~\eqref{p2}, hence the case for $L \in (L_{min},L_{Max})$ is contained in \Cref{thm:conv}.

\subsection{Proof of Lemma~\ref{thm:conv}}
\label{subsec:pfthm:conv}

We first present several necessary lemmas, and then give the proof of Lemma~\ref{thm:conv}. 

The partial minimization for $\bm{q}$ depends on $s$, the following corollary of Lemma~\ref{lem:Fs1} describes the behavior of the Lagrange function with parameter $s$ while doing the minimization for a different parameter $s'$. 
\begin{corollary}
\label{cor:Fs2}
\begin{equation}
\begin{aligned}
    F_s(\bm{q},\bm{r}) = F_s(\bm{q}_{s'}^*(\bm{r}),\bm{r})+ GD_{\mathcal{E}}(\bm{q}||\bm{q}_{s'}^*(\bm{r})) \\
    +(s-s')\left[Loss(\bm{q})-Loss(\bm{q}_{s'}^*(\bm{r}))\right].
\end{aligned}
\end{equation}
\end{corollary}
\begin{IEEEproof}
By the identity $F_s(\bm{q},\bm{r}) = F_{s'}(\bm{q},\bm{r})+(s-s')Loss(\bm{q})$ and \eqref{eq:Fsq} in \Cref{lem:Fs1}.
\end{IEEEproof}

\begin{corollary}
\label{cor:q0r0}
The conditional distribution pair $(\bm{q}^0,\bm{r}^0)$ minimizing the Lagrange function $F_{s^*}(\bm{q},\bm{r})$ exists and satisfies $\bm{q}_{s^*}^*(\bm{r}^0) = \bm{q}^0$. Also, $r^*(\bm{q}^0)(\cdot|w) = r^0(\cdot|w)$ for any $w$ such that $q^0(w) >0$.
\end{corollary}
\begin{IEEEproof}
The domain of $(\bm{q},\bm{r})$ is compact and $F_{s^*}(\bm{q},\bm{r})$ is continuous, so there exists some $(\bm{q}^0,\bm{r}^0)$ minimizing the Lagrange function $F_{s^*}(\bm{q},\bm{r})$.

Then by the definition of $(\bm{q}^0,\bm{r}^0)$, 
$$F_{s^*}(\bm{q}^0,\bm{r}^0) \leq F_{s^*}(\bm{q}_{s^*}^*(\bm{r}^0),\bm{r}^0),$$ 
$$F_{s^*}(\bm{q}^0,\bm{r}^0) \leq F_{s^*}(\bm{q}^0,\bm{r}^*(\bm{q}^0)).$$
So by \eqref{eq:Fsq} \Cref{lem:Fs1} we have $GD_{\mathcal{E}}(\bm{q}^0||\bm{q}_{s^*}^*(\bm{r}^0))=0$ and $GD(\bm{r}^*(\bm{q}^0)||\bm{r}^0)=0$.

By \eqref{positivepv} we always have $ \bm{q}_{s^*}^*(\bm{r}^0)=\bm{q}^0$. Similarly, $r^*(\bm{q}^0)(\cdot|w) = r^0(\cdot|w)$ for any $w$ such that $q^0(w) >0$.
\end{IEEEproof}

\begin{lemma}
If $\Delta_n(\bm{q}^0) \geq 0$, then
\begin{equation}
\label{eq:mono1}
F_{s^*}(\bm{q}^{(n)},\bm{r}^{(n)}) \leq F_{s^*}(\bm{q}^{(n)},\bm{r}^{(n-1)}).
\end{equation}
If $s^{(n)} = s^*$, then we have 
\begin{equation}
\label{eq:mono2}
F_{s^*}(\bm{q}^{(n+1)},\bm{r}^{(n)}) \leq F_{s^*}(\bm{q}^{(n)},\bm{r}^{(n)}) \leq F_{s^*}(\bm{q}^{(n)},\bm{r}^{(n-1)}).
\end{equation}
\end{lemma}
\begin{IEEEproof}
By \eqref{eq:Fsr} in \Cref{lem:Fs1}, we have 
\begin{align*}
&F_{s^*}(\bm{q}^{(n)},\bm{r}^{(n-1)})
\\
= &F_{s^*}(\bm{q}^{(n)},\bm{r}^*(\bm{q}^{(n)}))+ GD_{\bm{q}^{(n)}}(\bm{r}^*(\bm{q}^{(n)})||\bm{r}^{(n-1)})\\
=&F_{s^*}(\bm{q}^{(n)},\bm{r}^{(n)})+ GD_{\bm{q}^{(n)}}(\bm{r}^{(n)}||\bm{r}^{(n-1)}) 
\\
\geq& F_{s^*}(\bm{q}^{(n)},\bm{r}^{(n)}).
\end{align*}
Furthermore, if $s^{(n)} = s^*$, then by \Cref{cor:Fs2} and $\bm{q}_{s^{(n)}}^*(\bm{r}^{(n)}) = \bm{q}^{(n+1)}$,
\begin{align*}
&F_{s^*}(\bm{q}^{(n)},\bm{r}^{(n)})
= F_{s^*}(\bm{q}^{(n+1)},\bm{r}^{(n)})+GD_{\mathcal{E}}(\bm{q}^{(n)}||\bm{q}^{(n+1)})
\\
+ &(s^*-s^{(n)})(Loss(\bm{q}^{(n)})-Loss(\bm{q}^{(n+1)}))
\\
=&F_{s^*}(\bm{q}^{(n+1)},\bm{r}^{(n)})+GD_{\mathcal{E}}(\bm{q}^{(n)}||\bm{q}^{(n+1)}) \\
\geq
&F_{s^*}(\bm{q}^{(n+1)},\bm{r}^{(n)}).
\end{align*}
\end{IEEEproof}

\begin{lemma}
\label{lem:keyidentity}
For some $(\bm{q}^0,\bm{r}^0) \in \arg \min_{\bm{q},\bm{r}}F_{s^*}(\bm{q},\bm{r})$, define 
\begin{align*}
\Gamma_n(\bm{q}^0) = GD_{\mathcal{E}}(\bm{q}^0||\bm{q}^{(n)}) -GD_{\bm{q}^0}(\bm{r}^0||\bm{r}^{(n)}),
\end{align*}
then $\Gamma_n(\bm{q}^0) \geq 0$. Also, we have 
\begin{equation}
\begin{aligned}
&F_{s^*}(\bm{q}^{(n)},\bm{r}^{(n-1)})-F_{s^*}(\bm{q}^{0},\bm{r}^{0}) 
+\Gamma_{n-1}(\bm{q}^0)+ \Delta_n(\bm{q}^0)
\\
= &GD_{\mathcal{E}}(\bm{q}^{0}||\bm{q}^{(n-1)})-GD_{\mathcal{E}}(\bm{q}^{0}||\bm{q}^{(n)}).
\end{aligned}
\end{equation}
\end{lemma}
\begin{IEEEproof}
By the definition of $\Gamma_n(\bm{q}^0)$ $\bm{r}^{(n)} = \bm{r}^*(\bm{q}^{(n)})$ and~\eqref{eq:alterr}, we have
\begin{align*}
    &\ \Gamma_n(\bm{q}^0) 
    \\
    &= -\sum_{v,u,w}p(v)q^0(u|v)p(w|u,v)\log \frac{q^{(n)}(u|v)r^{0}(u|w)}{q^0(u|v)r^{(n)}(u|w)}
    \\
    &\geq -\sum_{v,u,w}p(v)q^0(u|v)p(w|u,v) \frac{q^{(n)}(u|v)r^{0}(u|w)}{q^0(u|v)r^{(n)}(u|w)}+1
    \\
    &= -\sum_{u,w}\sum_{v}p(v)p(w|u,v) \frac{q^{(n)}(u|v)r^{0}(u|w)}{r^{(n)}(u|w)}+1
    \\
    &= -\sum_{u,w}r^{0}(u|w)\sum_{u',v'}p(v')p(w|u',v') q^{(n)}(u'|v')+1
    \\
    &=0.
\end{align*}
Since $Loss(\bm{q}^{(n)}) = G_{\bm{r}^{(n-1)}}(s^{(n)})$, by \Cref{cor:Fs2} we have
\begin{equation}
\label{eq:lemmaFsq}
\begin{aligned}
F_{s^*}(\bm{q}^{0},\bm{r}^{(n-1)}) = F_{s^*}(\bm{q}^{(n)},\bm{r}^{(n-1)})+GD_{\mathcal{E}}(\bm{q}^0||\bm{q}^{(n)})\\
+(s^{(n)}-s^*)(G_{\bm{r}^{(n-1)}}(s^{(n)})-Loss(\bm{q}^0)).
\end{aligned}
\end{equation}
Note that by \Cref{cor:q0r0}, $\bm{r}^0 = \bm{r}^*(\bm{q}^0)$. Then by \eqref{eq:Fsr} in \Cref{lem:Fs1} we have
\begin{align}
\label{eq:lemmaFsr}
F_{s^*}(\bm{q}^{0},\bm{r}^{(n-1)}) = F_{s^*}(\bm{q}^{0},\bm{r}^{0})+GD_{\bm{q}^0}(\bm{r}^0||\bm{r}^{(n-1)}).
\end{align}

By \eqref{eq:lemmaFsq} and \eqref{eq:lemmaFsr} we have finished the proof.
\end{IEEEproof}

\textbf{Completing the Proof of \Cref{thm:conv}: }
Consider the first part and suppose $(\bm{q}^0,\bm{r}^0) \in \arg \min_{\bm{q},\bm{r}}F_{s^*}(\bm{q},\bm{r})$. By \Cref{lem:keyidentity}, we take the sum and get  
\begin{equation}\label{dif}
\begin{aligned}
&\sum_{k = n+1}^{m}\!\!\!\!F_{s^*}(\bm{q}^{(k)}\!,\!\bm{r}^{(k-1)})\!-\!F_{s^*}(\bm{q}^{0},\bm{r}^{0})\!+\!\Gamma_{k-1}(\bm{q}^0)\!+\!\Delta_k(\bm{q}^0)
\\
&=GD_{\mathcal{E}}(\bm{q}^0||\bm{q}^{(n)})-GD_{\mathcal{E}}(\bm{q}^0||\bm{q}^{(m)}),
\end{aligned}
\end{equation}
for $m > n \geq 1$.
Take $n=1$ and by the non-negativity of $\Gamma_{k-1}(\bm{q}^0)$ and $\Delta_k(\bm{q}^0)$, we can obtain
\[
\begin{split}
&\sum_{k = 2}^{m}\left(F_{s^*}(\bm{q}^{(k)},\bm{r}^{(k-1)})-F_{s^*}(\bm{q}^{0},\bm{r}^{0}) \right)
\\
\leq &GD_{\mathcal{E}}(\bm{q}^0||\bm{q}^{(1)}) \leq \log|\mathcal{U}|,
\end{split}
\]
which is because $q^{(1)}(u|v) = \frac{\mathds{1}_{(u \in \mathcal{E}_v)}}{|\mathcal{E}_v|}$.
Then~\eqref{eq:generalrate} is immediately implied.
Let $m \to \infty$, then we have  
\begin{equation*}
\sum_{k = 2}^{\infty}\left(F_{s^*}(\bm{q}^{(k)},\bm{r}^{(k-1)})-F_{s^*}(\bm{q}^{0},\bm{r}^{0})\right) \leq \log|\mathcal{U}|.
\end{equation*}
Each term in the sum is nonnegative, so $$\lim_{k \to \infty}F_{s^*}(\bm{q}^{(k)},\bm{r}^{(k-1)})=F_{s^*}(\bm{q}^0,\bm{r}^0).$$
Also, we have $\lim_{k \to \infty}F_{s^*}(\bm{q}^{(k)},\bm{r}^{(k)})=F_{s^*}(\bm{q}^0,\bm{r}^0) = \min_{\bm{q},\bm{r}}F_{s^*}(\bm{q},\bm{r})$ by \eqref{eq:mono1}, which proves the first part.

Then we consider the second part.
For~\eqref{pallc}, let $s^*$ be the Lagrange multiplier  satisfying the optimality condition.
Then by analyzing the KKT conditions, any optimal solution $(\bm{q}^0,\bm{r}^0)$ of (P) satisfies $(\bm{q}^0,\bm{r}^0) \in \arg \min_{\bm{q},\bm{r}}F_{s^*}(\bm{q},\bm{r})$.
$\{\bm{q}^{(n)}\}_{n \geq 1}$ is a sequence in a compact set, and hence has a convergent subsequence, donoted by $\{\bm{q}^{(n_k)}\}_{k=1}^{\infty}$. Let $\bm{q}^{(0)}$ be its limit, and let $\bm{r}^{(0)} = \bm{r}^*(\bm{q}^{(0)})$.
Then $\bm{r}^{(n_k) }= \bm{r}^*(\bm{q}^{(n_k)})$ and 
\begin{align*}
&\lim_{k \to \infty} F_{s^*}(\bm{q}^{(n_k)},\bm{r}^{(n_k)}) 
=\lim_{k \to \infty} F_{s^*}(\bm{q}^{(n_k)},\bm{r}^*(\bm{q}^{(n_k)}))\\
=&F_{s^*}(\bm{q}^{(0)},\bm{r}^*(\bm{q}^{(0)})) =  F_{s^*}(\bm{q}^{(0)},\bm{r}^{(0)}),
\end{align*}
which implies $F_{s^*}(\bm{q}^{(0)},\bm{r}^{(0)}) = \min_{\bm{q},\bm{r}}F_{s^*}(\bm{q},\bm{r})$.
Also, we have
\begin{align*}
    Loss(\bm{q}^{(0)}) = \lim_{k \to \infty}Loss(\bm{q}^{(n_k)}),
\end{align*}
and hence $\bm{q}^{(0)}$ satisfies the constraint in~\eqref{pallc}.
So $(\bm{q}^{(0)},\bm{r}^{(0)})$  is an optimal solution for the corresponding problem, which implies that 
 \eqref{dif} is also satisfied when $(\bm{q}^0,\bm{r}^0)$ is replaced by $(\bm{q}^{(0)},\bm{r}^{(0)})$.

Now let (P) be~\eqref{palla} or~\eqref{pallc}. By the version of \eqref{dif} where $(\bm{q}^0,\bm{r}^0)$ is replaced by $(\bm{q}^{(0)},\bm{r}^{(0)})$, we have $GD_{\mathcal{E}}(\bm{q}^{(0)}||\bm{q}^{(n)})$ is non-increasing. Since  $\lim_{k \to \infty}\bm{q}^{(n_k)} = \bm{q}^{(0)}$, then we have $\lim_{k \to \infty}GD_{\mathcal{E}}(\bm{q}^{(0)}||\bm{q}^{(n_k)}) =0$ and hence $\lim_{n \to \infty}GD_{\mathcal{E}}(\bm{q}^{(0)}||\bm{q}^{(n)}) =0$. So $\bm{q}^{(n)} \to \bm{q}^{(0)}$, which implies $\bm{r}^{(n)} \to \bm{r}^{(0)}$ as $n \to \infty$.
In other words, the solutions $(\bm{q}^{(n+1)}, \bm{r}^{(n)})$ converge to an optimal solution $(\bm{q}^{(0)}, \bm{r}^{(0)})$ for the corresponding
problem~\eqref{palla} or~\eqref{pallc}.

For the case $s^{(n)} = s^*$, by \eqref{eq:mono2} $F_{s^*}(\bm{q}^{(n+1)},\bm{r}^{(n)}) - F_{s^*}(\bm{q}^0,\bm{r}^0)$ is non-increasing, hence by~\eqref{eq:generalrate} it is no greater than $\frac{\log |\mathcal{U}|}{n}$, which completes the proof.

\subsection{Proof of Lemma~\ref{lem:greaterKKT}}
\label{subsec:pflem:greaterKKT}
Analyzing KKT conditions of the problem \eqref{pallc} for $L > L_{min}$, there are two cases.

\begin{enumerate}
    \item For $L\leq L_{max}$, we have $s^* \geq 0$ and $Loss(\bm{q}^0) = L$.
    Then $ \Delta_n(\bm{q}^0) = (s^{(n)}-s^*)(G_{\bm{r}^{(n-1)}}(s^{(n)})-L)$.
    For case i), $s^{(n)} = 0 \leq s^*$ and  $G_{\bm{r}^{(n-1)}}(s^{(n)}) = G_{\bm{r}^{(n-1)}}(0)\leq L$, so $ \Delta_n(\bm{q}^0) \geq 0$. For case ii), $G_{\bm{r}^{(n-1)}}(s^{(n)})=L$ and hence $ \Delta_n(\bm{q}^0) = 0$.

    \item For $L > L_{max}$, we have $s^* = 0$ and $Loss(\bm{q}^0) \leq L$.  For case i), $s^{(n)} = 0 = s^*$ and hence $ \Delta_n(\bm{q}^0) = 0$. For case ii), $s^{(n)} \geq 0 = s^*$ and $G_{\bm{r}^{(n-1)}}(s^{(n)}) = L \geq Loss(\bm{q}^0)$, so $ \Delta_n(\bm{q}^0) \geq 0$.
\end{enumerate}

\section{Proof of Lemma \ref{lem:sparsity}}
\label{sec:pfspar}
Let $U$ be optimal with joint distribution $p(v,u,w) = p(u)p(v,w|u)$ for~\eqref{p0}. 
Note that 
\begin{align*}
    &H(W|U)-H(V|U) \\
    &= \sum_u p(u)(H(W|U=u)-H(V|U=u)),
    \\
    & \mathbb{E}[l(V,U,W)] = \sum_u p(u)\sum_{v,w}p(v,w|u)l(v,u,w),
    \\
    &p(v) = \sum_u p(u) p(v|u),v\in \mathcal{V}-\{v_0\},
    \\
    &p(w) = \sum_u p(u) p(w|u), w \in \mathcal{W}-\{w_0\},
\end{align*}
where $v_0 \in \mathcal{V}$ and $w_0 \in \mathcal{W}$.
There are totally
$1+1+(|\mathcal{V}|-1)+(|\mathcal{W}|-1) = |\mathcal{V}|+|\mathcal{W}|
$
equations. By the Fenchel–Eggleston–Carath\'{e}odory theorem (or the support lemma in Appendix~C of~\cite{EIGamal2011}), there exists some $U'$ with alphabet $\mathcal{U}$ and joint distribution $p'(u,v,w)= p'(u)p(v,w|u)$ such that i) the above equations hold if $p(u)$ is replaced by $p'(u)$ on the right hand side;
ii) there are at most $|\mathcal{V}|+|\mathcal{W}|$ of $u$ which satisfy $p'(u)>0$.
Note that
\begin{equation*}
\begin{aligned}
    &I(U;V) - I(U;W) 
    \\
    = &H(V)-H(W)-H(V|U)+H(W|U).
\end{aligned}
\end{equation*}
Then by i), we have $I(U';V)-I(U';W) = I(U;V) - I(U;W)$ which shows the optimality of $U'$ and completes the proof.






\end{document}